\documentclass[11pt]{article}
\usepackage[letterpaper]{geometry}

\usepackage{fullpage}
\usepackage[utf8]{inputenc}
\usepackage{amssymb, amsthm, amsfonts}
\usepackage{setspace}
\usepackage{fancyhdr}
\usepackage{lastpage}
\usepackage{extramarks}
\usepackage{chngpage}
\usepackage{soul,color}
\usepackage{graphicx,float,wrapfig}
\usepackage{listings}
\usepackage{xcolor}      
\usepackage{enumerate}
\usepackage{enumitem}
\usepackage{xspace}
\usepackage[T1]{fontenc}
\usepackage{subfiles}
\usepackage{bm}
\usepackage{tabularx}
\usepackage{multirow}
\usepackage{tcolorbox}

\usepackage{xcolor}
\usepackage{nameref}
\usepackage{fdsymbol}
\definecolor{ForestGreen}{rgb}{0.1333,0.5451,0.1333}
\definecolor{DarkRed}{rgb}{0.65,0,0}
\definecolor{Red}{rgb}{1,0,0}
\usepackage[linktocpage=true,
pagebackref=true,colorlinks,
linkcolor=DarkRed,citecolor=ForestGreen,
bookmarks,bookmarksopen,bookmarksnumbered]
{hyperref}
\usepackage{cleveref}
\usepackage{amsthm}
\usepackage{algpseudocode}
\usepackage{algorithm}  
\usepackage{algorithmicx} 
\usepackage{verbatim}
\usepackage{todonotes}
\usepackage{appendix}

\makeatletter
\def\namedlabel#1#2{\begingroup
    #2%
    \def\@currentlabel{#2}%
    \phantomsection\label{#1}\endgroup
}
\makeatother

\usepackage{thmtools}
\usepackage{thm-restate}

\declaretheorem[numberwithin=section]{theorem}
\declaretheorem[numberlike=theorem]{lemma}

\declaretheorem[numberlike=theorem]{corollary}

\declaretheorem[numberlike=theorem,style=definition]{definition}

\declaretheorem[numberlike=theorem,style=definition]{remark}

\usepackage{mathrsfs}

\newcommand{\polylog}{\operatorname{\text{{\rm polylog}}}}

\newcommand{\ID}{\operatorname{\text{{\rm ID}}}}

\newcommand{\OPT}{\mathrm{OPT}}

\newcommand{\obj}{\mathrm{obj}}

\newcommand{\cand}{\mathrm{Cand}}

\newcommand{\Vhigh}{V_{\operatorname{high}}}
\newcommand{\Vlow}{V_{\operatorname{low}}}
\newcommand{\Gsim}{G_{\operatorname{sim}}}

\newcommand{\Esim}{E_{\operatorname{sim}}}
\renewcommand{\deg}{d}

\Crefname{ALC@unique}{Line}{Lines}
\Crefname{@theorem}{Theorem}{Theorems}

\begin{document}

\newcommand\relatedversion{}
\renewcommand\relatedversion{\thanks{This work is supported by NSF CCF-2008422.}} 

\title{\Large Min-Max Correlation Clustering via Neighborhood Similarity}
\author{Nairen Cao\\
New York University\and 
Steven Roche\\
Boston College\and
Hsin-Hao Su \thanks{Supported by NSF CCF-2008422.}\\
Boston College 
}

\date{}

\maketitle


\begin{abstract}\small 
We present an efficient algorithm for the min-max correlation clustering problem. The input is a complete graph where edges are labeled as either positive $(+)$ or negative $(-)$, and the objective is to find a clustering that minimizes the $\ell_{\infty}$-norm of the disagreement vector over all vertices.

We resolve this problem with an efficient $(3 + \epsilon)$-approximation algorithm that runs in nearly linear time, $\tilde{O}(|E^+|)$, where $|E^+|$ denotes the number of positive edges. This improves upon the previous best-known approximation guarantee of 4 by Heidrich, Irmai, and
Andres~\cite{heidrich20244}, whose algorithm runs in $O(|V|^2 + |V| D^2)$ time, where $|V|$ is the number of nodes and $D$ is the maximum degree in the graph.

Furthermore, we extend our algorithm to the massively parallel computation (MPC) model and the semi-streaming model. In the MPC model, our algorithm runs on machines with memory sublinear in the number of nodes and takes $O(1)$ rounds. In the streaming model, our algorithm requires only $\tilde{O}(|V|)$ space, where $|V|$ is the number of vertices in the graph.

Our algorithms are purely combinatorial. They are based on a novel structural observation about the optimal min-max instance, which enables the construction of a $(3 + \epsilon)$-approximation algorithm using $O(|E^+|)$ neighborhood similarity queries. By leveraging random projection, we further show these queries can be computed in nearly linear time.


\end{abstract}
\sloppy
\section{Introduction}
In the correlation clustering problem, we are given a complete graph where each edge is labeled as either \( "+" \) or \( "-" \). A \( "+" \) edge indicates that the two vertices are \textit{similar}, while a \( "-" \) edge indicates they are \textit{dissimilar}. For any partition of the graph, an edge is considered to be in disagreement if it is a negative edge and its endpoints belong to the same cluster, or if it is a positive edge and its endpoints belong to different clusters.

Given a clustering (partition), the disagreement \( \rho(v) \) for any node \( v \) is defined as the number of edges incident to \( v \) that are in disagreement with respect to the clustering. The goal of the correlation clustering problem is to find a clustering that minimizes an objective function capturing the disagreement of edges.

Puleo and Milenkovic~\cite{PM18} introduced the objective of minimizing the \( \ell_p \)-norm of the disagreements over the vertices, which is defined as:
\[
\bigg( \sum_{v \in V} \rho(v)^p \bigg)^{1/p}.
\]
This objective generalizes the correlation clustering problem proposed by Bansal, Blum, and Chawla~\cite{BBC04}, which corresponds to the case \( p = 1 \), where the goal is to minimize the total number of disagreements. Significant progress has been made on the \( \ell_1 \)-norm objective~\cite{CGW05,ACN08,CMSY15,CLN22,CLLN23}, Cao, Cohen-Addad, Lee,
Li, Newman, and Vogl~\cite{cao2024understanding} present a $1.437$-approximate algorithm for \( \ell_1 \)-norm objective. For the case \( p = \infty \), it corresponds to the min-max correlation clustering problem, where the goal is to minimize the maximum disagreements over each vertex. The problem remains relatively unexplored and the best approximate ratio is $4$ given by Heidrich, Irmai, and Andres~\cite{heidrich20244}.

Puleo and Milenkovic~\cite{PM18} proposed an algorithm that achieves a $48$-approximation ratio. Their approach uses the standard metric linear programming formulation followed by a rounding algorithm. Later, Charikar, Gupta, and Schwartz~\cite{CGS17} improved this result to a $7$-approximation using the same framework, and Kalhan, Makarychev, and Zhou~\cite{KMZ19} further reduced it to a $5$-approximation. Davies, Moseley, and Newman~\cite{davies2023fast} designed a combinatorial algorithm that achieves a $40$-approximation ratio with a runtime of $O(n^2 \log n)$, where $n = |V|$ is the number of vertices in the graph. The best known approximation ratio to date is $4$, achieved by Heidrich, Irmai, and Andres~\cite{heidrich20244}, who also used a combinatorial approach with a runtime of $O(n^2 + n D^2)$, where $D$ is the maximum degree of the graph.

Concerning efficiency, the ultimate goal of an efficient algorithm is to achieve a running time that is nearly linear in $m = |E^{+}|$.\footnote{In previous correlation clustering literature, it is typical to use $m = |E^{+}|$ to denote the number of positive edges and to obtain bounds in terms of $m$, as it has been pointed out by \cite{CDK14} that it is common to have a much smaller number of positive edges than negative edges in practical applications.} However, none of the aforementioned algorithms has yet achieved such a goal. Recently, Cao, Li, and Ye~\cite{cao2024simultaneously} proposed an nearly-linear time algorithm that achieves a $63.3$-approximation ratio. While their algorithm is fast, the approximation ratio is far from optimal.


This naturally leads to the question:

\begin{quote}
    Can we design a nearly linear algorithm for min-max correlation clustering with a small approximation ratio?
\end{quote}

We give a nearly linear algorithm for the problem that achieves a $(3+\epsilon)$-approximation. The algorithm can be tailored into algorithms in various large-scale models, such as a $O(1)$-round algorithm for the sublinear-memory massively parallel computation (MPC) model, as well as a single-pass semi-streaming algorithm. Our main result is given as follows:

\begin{restatable}{theorem}{thmmain} \label{thm:thmmain}
Let $G = (V, E^+)$ be a min-max correlation clustering instance, $\epsilon > 0$ be a small constant, and $\OPT$ be the value of the optimal solution. In the following models, there exist randomized algorithms that output a clustering $\mathcal{C}$ with $\obj(\mathcal{C}) \leq (3+\epsilon)\cdot \OPT$ w.h.p.\footnote{{\it With high probability}, which refers to with probability at least $1-1/n^{c}$ for some sufficiently large constant $c$.} with the following attributes:
\begin{enumerate}
    \item \label{itm:main1} (Sequential model) An $O(m \log^2 n/ \epsilon^2 )$-time algorithm.
    
    
    \item \label{itm:main2}(MPC model) An $O(\log(1/\epsilon))$-round algorithm using $O(n^{\delta})$ memory per machine and total memory $O(m \log n / \epsilon^2)$.
    
    \item \label{itm:main3}(Semi-streaming model) A single-pass streaming algorithms that uses $O(n \log n/ \epsilon^2)$ space.
\end{enumerate}
\end{restatable}

We may also trade the running time for an exact 3-factor approximation algorithm. 
\begin{restatable}{cor}{exactthreeapprox}\label{cor:exact3approx}
There exists a deterministic sequential algorithm that runs in $O(m (D\log D) (\log n))$ time that outputs a 3-approximate solution, where $D$ is the maximum degree of $G^{+} = (V, E^{+})$. 
\end{restatable}


\subsection{Additional Related Works}

A substantial body of research has focused on parallel algorithms and streaming algorithms for the correlation clustering problem. In the MPC model, significantly more work has been done for the $\ell_1$ setting. Specifically, numerous studies have focused on improving the number of rounds and the approximation ratio~\cite{blelloch2012greedy, PPORRJ15, FischerN20, CCMU21, DBLP:conf/icml/Cohen-AddadLMNP21, assadi2021sublinear, CKLPU23, cao2024breaking}. Very recently, Cohen-Addad, Lolck, Pilipczuk, Thorup, Yan, and Zhang~\cite{cohen2024combinatorial} designed an $O(1)$-round MPC algorithm that achieves a 1.876-approximation ratio.

In contrast, there has been less work for the $\ell_{\infty}$ setting. The only known work is by Cao, Ye, and Li~\cite{cao2024simultaneously}, who proposed an algorithm that achieves a 63.3-approximation ratio in $O(\log^3 n)$ rounds and another algorithm that achieves a 360-approximation ratio in $O(1)$ rounds.

The correlation clustering problem has also been extensively studied in the streaming model for the $\ell_1$ setting. Several algorithms have achieved good approximation ratios with constant rounds~\cite{CDK14, ACGMW21, DBLP:conf/icml/Cohen-AddadLMNP21, BCMT22}. In the single-pass setting, Ahn, Cormode, Guha, McGregor, and Wirth~\cite{ACGMW21} presented the first algorithm with a space complexity of $\tilde{O}(n + m^-)$, where $m^-$ is the number of negative edges in the graph. This result has been further improved by multiple works~\cite{assadi2021sublinear, BCMT22, chakrabarty2023single}. Very recently, Cohen-Addad, Lolck, Pilipczuk, Thorup, Yan, and Zhang~\cite{cohen2024combinatorial} designed an algorithm that achieves a 1.876-approximation ratio in a single pass using $n \polylog(n)$ space.

There has been very little work explicitly addressing the $\ell_{\infty}$ case in the streaming model. To the best of our knowledge, we are the first to propose a streaming algorithm for the $\ell_{\infty}$ correlation clustering problem.

\subsection{Technical Overview} 
\paragraph{Better Approximation} Our first main technical contribution is a newly achieved approximation factor of 3. Given a guess for the optimal objective value $\phi$, if $OPT \leq \phi$, \cite{heidrich20244} observed that if the neighborhoods of $u$ and $v$ share at least $2\phi$ elements, then they must belong to the same cluster in the optimal solution. Similarly, if $u$ and $v$ differ by more than $2\phi$ elements, then they must belong to different clusters in the optimal solution. 

Furthermore, they observed that these properties can be used to determine the clusters for vertices of degrees at least $4\phi$. Specifically, if $\deg(x) \geq 4\phi$, then for every other vertex $y$, either $|N[x] \cap N[y]| \geq 2\phi$ or $|N[x] \Delta N[y]| \geq 2\phi$. Here, $N[x] = N(x) \cup \{x\}$ represents the closed neighborhood of vertex $x$. The remaining vertices can then be placed in singleton clusters, as the disagreements per vertex will be upper bounded by their degrees, $4\phi$. Therefore, a clustering of disagreements upper bounded by $4\phi$ can be constructed, resulting in a 4-approximation algorithm. 

To achieve a 3-approximation, we first observe that if two vertices $x$ and $y$ have degrees greater than $3\phi$, then it is also the case either $|N[x] \cap N[y]| \geq 2\phi$ or $|N[x] \Delta N[y]| \geq 2\phi$ holds. In other words, whether $x$ and $y$ belong to the same cluster is uniquely determined in the optimal solution. Therefore, the clustering induced on the high-degree vertices (vertices with $\deg > 3\phi$) is uniquely determined.

Now the question lies in the placement of the low-degree vertices, that is, vertices with degrees upper-bounded by $3\phi$. It is unclear whether they should be placed in singleton clusters, as it is possible that they need to be included in the same cluster with certain high-degree vertices. Otherwise, the disagreements associated with the high-degree vertices could become too large.

We show that the low-degree vertices can be placed in the high-degree clusters to achieve a maximum disagreements of $3\phi$, provided $OPT \leq \phi$. A key structural result we show is that if a low-degree vertex $w$ belongs to some cluster $C$ in an optimal solution, then no vertex $v$ outside $C$ can have similar neighborhood with $w$, i.e., $|N[v] \Delta N[w]| \leq 2\phi$. Using this structural result, we show the following algorithm constructs a clustering with maximum disagreement upper bounded by $3\phi$ (presented slightly differently here than in the main body for the sake of intuition):

1. Form clusters for high-degree vertices based on whether $|N[u] \Delta N[v]| \leq 2\phi$ for all high-degree vertices $u,v$. 2. Choose an arbitrary vertex $u$ in each cluster and have it propose to low-degree neighboring vertices with whose neighborhoods are similar to $u$. 3. For each low-degree vertex that receives at least one proposal, pick one arbitrary proposal and join the cluster containing the vertex that sent it. 4. Place all low-degree vertices that do not receive a proposal into singleton clusters.

\paragraph{Efficient Implementations} 
The remaining question is how such an algorithm can be implemented efficiently, particularly in time (and total memory) nearly linear in $|E^{+}|$. A main technical challenge lies in Step 1. To implement it within the aforementioned time bound, we can only afford to conduct similarity tests (i.e.~to test whether $|N[u] \Delta N[v]| \leq 2\phi$) for $O(|E^{+}|)$ times. However, it is possible for two vertices that are endpoints of a negative edge to have similar neighborhoods and thus need to be placed in the same cluster to achieve a good clustering.

Using the structural result, we further show that two high-degree vertices $u$ and $v$ are in the same cluster in the optimal solution if and only if there are at least $\phi+1$ disjoint paths of length 2 connecting $u$ and $v$ in $\Esim$, where $\Esim \subseteq E^{+}$ consists of all the edges in $E^{+}$ whose endpoints have similar neighborhoods. This property enables us to develop efficient algorithms for the sublinear MPC model and the sequential model.

The remaining question lies in how to find $\Esim$ efficiently. To this end, for each vertex $u$, we treat its neighborhood set $N[u]$ as a point in an $n$-dimensional space. Then, we apply the (discrete) random projection technique \cite{Achlioptas03} developed for the Johnson-Lindenstrauss transform \cite{JL84} to reduce the dimension to $O(\log n / \epsilon^2)$ while preserving the $\ell_2$ distance (up to a $(1 \pm \epsilon)$ factor) between the points. For 0/1 vectors, the $\ell_2$ distance is exactly the symmetric difference. Since the dimension is $O(\log n / \epsilon^2)$, it takes $O(\log n / \epsilon^2)$ time to compute the difference. To our knowledge, this is the first time that random projection techniques have been applied to computing efficient solutions for correlation clustering and problems alike. This may be of independent interest, as neighborhood similarity is known to be used in various tools such as {\it almost-clique decompositions} \cite{HSS18, 
CLP18, HKMT21, HKNT22, FGHKN24, ACK19, FHM23, AKM23, CLMNP21,AW22}.

\paragraph{Single-Pass Semi-Streaming} While the above techniques are sufficient for getting our  MPC and sequential algorithms, the single-pass semi-streaming algorithm introduces additional technical difficulties. The main difficulty for a single-pass semi-streaming to work here lies in Step 2, where an arbitrary chosen high-degree vertex in each cluster proposes to its neighbors who have similar neighborhoods. For convenience, we call those vertices {\it pivots} here. 

To be able to do this, we need to memorize the neighbors of all the chosen vertices using $n \polylog (n)$ space. Suppose that $\OPT \leq \phi$, it can be shown that each cluster in the optimal clustering containing at least one low-degree vertex has size of $\Theta(\phi)$. As a result, any vertex from these clusters has degree at most $O(\phi)$. Since we pick a pivot per cluster, $O(\phi \cdot (n/\phi))$ is the space we need to store the neighbors of the pivots. 

However, we do not know beforehand how the clusters of high-degree vertices look like, so the pivots cannot be chosen at the beginning of the stream. Without knowing what the pivots are beforehand, it is difficult to store their neighbors in the same pass. 

As a result, we sample each vertex (both high-degree and low-degree vertices) independently with probability $O(\log n/ d(v))$ so that w.h.p.~each cluster in the optimal clustering has $O(\log n)$ sampled vertices. Then we store the neighbors of all the sampled vertices. This poses another problem: Low-degree vertices may be chosen as pivots. However, a low-degree vertex is exempted from our structural result -- using it as a pivot may steal vertices from other clusters in an optimal clustering. 

To resolve this, we do the following. For each sampled vertex $y$, we first try to recover the high-degree portion $L$ of the cluster containing $y$ in the optimal solution. We construct a candidate set $\cand(L)$ that contains vertices that would not be added to other clusters. When using $y$ as a pivot, we restrict it to consider only the intersection with the candidate set, $N[y] \cap \cand(L)$ to ensure that it does not steal vertices from other clusters. 

Roughly speaking, the candidate set $\cand(L)$ contains all the low-degree vertices that have similar neighborhood with every vertex in $L$ but have different neighborhood with every vertex in any other cluster $L'$ (as a result, $\cand(L) \cap \cand(L')=\emptyset$). We show such a modification does not affect the approximation ratio. Furthermore, since the candidate sets are defined based on similarity of neighborhoods, they can be constructed by the aforementioned dimension reduction technique, which takes $O(n \log n /\epsilon^2)$ space. 



\section{Preliminaries}
\begin{definition}
Given $u \in G^{+} = (V,E^{+})$, let $N(u)$ denote the neighbors of $u \in G^{+}$. Define $N[u] = N(u) \cup \{u\}$. $d(u) = |N(u)|$ is the number of neighbors of $u \in G^+$. We use $N_{H}(u)$ ,$N_{H}[u]$ and $d_{H}(u)$ to denote the the corresponding quantities in graph $H$ when $H \neq G^{+}$.
\end{definition}

\begin{definition}
Let $A$ and $B$ be sets. The symmetry difference between $A$ and $B$, $A \Delta B$, is defined as $A \Delta B = (A \setminus B) \cup (B \setminus A)$.
\end{definition}
In general, we say $A$ and $B$ are similar if $|A \Delta B|$ is small.
\begin{lemma}[Triangle Inequality~\cite{halmos1960naive}]
\label{lem:triangleineq}
Let $A, B, C$ be sets. Then: $|A \Delta C| \le |A \Delta B| + |B \Delta C|$.
\end{lemma}

\begin{definition}
Given any clustering $\mathcal{C}$ and any vertex $u$, $\mathcal{C}_u$ is defined to be the cluster of $\mathcal{C}$ containing $u$. 
\end{definition}

\begin{definition}
Given clustering $\mathcal{C}$, define $\rho_{\mathcal{C}}(x)$ as the number of disagreements incident to vertex $x$. More precisely:
\begin{align*}
    \rho_{\mathcal{C}}(x) = |\mathcal{C}_x \setminus N[x]| + |N[x] \setminus \mathcal{C}_x| = |N[x] \Delta \mathcal{C}_x| = |N(x)~\Delta~\mathcal{C}_x| - 1
\end{align*}
\end{definition}

\begin{definition}
Given a clustering $\mathcal{C}$, the objective function of $\mathcal{C}$, $\obj(\mathcal{C}) = \max_{u} \rho_{\mathcal{C}}(u)$, is defined as the maximum incident disagreements over every vertex. 
\end{definition}

\paragraph{The MPC Model}
In the MPC model, computation proceeds in synchronous parallel \emph{rounds} across multiple machines. Each machine has memory $S$. At the beginning of the computation, data is arbitrarily partitioned across the machines. During each round, machines process data locally, exchange messages with other machines, and send or receive messages of total size $S$. The efficiency of an algorithm in this model is measured by the number of rounds required for the algorithm to terminate and the size $S$ of the memory available to each machine.

In this paper, we focus on the most practical and challenging regime, also known as the \emph{strictly sublinear regime}, where each machine has $S = O(n^\delta)$ local memory. Here, $n$ represents the input size, and $\delta$ is an arbitrary constant smaller than 1. Under this assumption, the input assigned to each machine and the messages exchanged during any round are of size $O(n^\delta)$.

\paragraph{The semi-streaming model.} In the semi-streaming model, the input is a stream of edges in \( E^{+} \). We are allowed to use \( n \polylog n \) space, where the space complexity refers to the number of words used. A solution is expected to be output at the end of the stream.

\section{Algorithm}

\begin{definition} An $\eta$-similarity query $\Delta_{\eta}(u,v,t)$ returns: 
$$\Delta_{\eta}(u,v,t) =   \begin{cases}0, &\mbox{if $|N[u] \Delta N[v]| > (1+\eta)t$.} \\ \mbox{$0$ or $1$}, &\mbox{if $t <  |N[u] \Delta N[v]| \leq (1+\eta)t$.} \\ 1, &\mbox{if $|N[u] \Delta N[v]| \leq t$.} \end{cases}$$.
\end{definition}

\begin{definition} (Shorthand for $\eta$-similarity query) For readability, we use \( |N[u] \Delta N[v]| \leq_{\eta} t \) to denote that an \( \eta \)-similarity query \( \Delta_{\eta}(u,v,t) \) has been conducted and returned 1. We use \( |N[u] \Delta N[v]| \nleq_{\eta} t \) to denote that the query \( \Delta_{\eta}(u,v,t) \) returned 0.
\end{definition}

The parameter $\eta$ was introduced to accommodate the error induced by an approximate test of whether $|N[u] \Delta N[v]| \leq 2\phi$. For the sake of simplicity, we urge the readers to assume $\eta = 0$ when reading for the first time. When $\eta = 0$, $|N[u] \Delta N[v]| \leq_{\eta} 2\phi$ if and only if $|N[u] \Delta N[v]| \leq 2\phi$.

\begin{definition} Define $\Vlow \leftarrow \{w \mid \mbox{$\deg(w) \leq (3+\eta) \phi$} \}$ and $\Vhigh \leftarrow V \setminus \Vlow$. \end{definition}


\paragraph{Algorithm Description} \Cref{alg:lp} takes two parameters $\phi$ and $0 \leq \eta < 1$, where $\phi$ is a guess on the upper bound of $\OPT$ and $\eta$ is an error control parameter.  The goal of the algorithm is to output a solution of value $(3+\eta)\phi$ provided $\OPT \leq \phi$. Note that we use $\eta$ instead of $\epsilon$ here to indicate that it can be set to zero (at the cost of computing the more expensive exact neighborhood similarity). 


The algorithm works as follows: First, we form a clustering of high-degree vertices $\mathcal{L}$ based on the similarity between the neighborhood of vertices. If two high-degree vertices $u$ and $v$ have similar neighborhood (i.e.~$N[u] \Delta N[v]| \leq_{\eta} 2\phi$) then they will be placed in the same cluster. 

Once the high-degree clusters are formed, we go through each cluster $L_i \in \mathcal{L}$. For each $L_i$, we pick a pivot $u_i \in L_i$. For each neighbor $w$ of $u_i$ that remains unclustered (i.e.~in $V_i$), we include it in $R(u_i)$ if (i.e.~$N[w] \Delta N[u_i]| \leq_{\eta} 2\phi$). Then we set our cluster $C_i$ to be $L_i \cup R(u_i)$. Then we update the unclustered vertices to be $V_{i+1} = V_{i} \setminus R(u_i)$. 

Once we went through every $L_i$, there might be still some unclustered low-degree vertices (i.e.~those in $V_{|\mathcal{L}|+1}$). For each such vertex, we put it in a singleton cluster. Then, we check if the clustering we have obtained has an objective value at most $(3+\eta)\phi$ or not. If it does, then we are done. If not, we conclude that $\OPT > \phi$, so we would need to set our guess of $\phi$ larger.

\begin{algorithm}[t]
\caption{{\textsc{ClusterPhi}}$(G^+=(V, E^+), \phi, \eta)$\label{alg:main}\\
\textbf{Input}: A graph $G^{+}$ and parameters $\phi$ and $0 \leq \eta < 1$. \\
\textbf{Output}: A clustering $\mathcal{C}$ with $\obj(\mathcal{C})\leq (3+\eta)\phi$ or ``$\OPT > \phi$''}
\label{alg:lp}
\begin{algorithmic}[1]
\Function {ClusterPhi}{$G^+=(V, E^+), \phi, \eta$}
\Statex {\small \textit{\underline{$\triangleright$ Initialization}}}


    \For{$u,v \in V$} \label{ln:highdegstart}\Comment{$uv$ not necessarily in $E^{+}$.}
        \If{$|N[u] \Delta N[v]| \leq_{\eta} 2 \phi$}
            \State $E' \leftarrow E' \cup \{uv\}$.
        \EndIf
     \EndFor
    
    \State $\Vlow \leftarrow \{w \mid \mbox{$\deg(w) \leq (3+\eta) \phi$} \}, \Vhigh \leftarrow V \setminus \Vlow$

    \State $V_1 = \Vlow$
    
    \State Let $\mathcal{L} = \{L_i\}_{i=1}^{|\mathcal{L}|}$ be the partition formed by the connected components in $(\Vhigh ,E')$. \label{ln:highdegend}

    \For{$i$ from $1$ to $|\mathcal{L}|$} \label{ln:upper} 
    \State Choose any node $u_i \in L_i$.
    \State \label{ln:R}Compute $R(u_i) = \{ w \in V_i \cap N(u_i) \mid |N[w] \Delta N[u_i]| \leq_{\eta/2} 2\phi   \}$
    \State  \label{ln:C}$C_i \leftarrow L_i \cup R(u_i)$                                                  
    \State $V_{i + 1} \leftarrow V_i \setminus R(u_i)$
    \EndFor   \label{ln:lower} 
    \If{for some $i$ there exists $u \in C_i$ such that $\rho_{C_{i}}(u) > (3+\eta)\phi$}\label{ln:checking}
    \State \Return ``$\OPT > \phi$'' 
    \Else
    \State \Return $\mathcal{C} = \{C_i\}_{i=1}^{|\mathcal{L}|} \cup \bigcup_{v \in V_{|\mathcal{L}|+1}}\{ \{v\} \}$
    \EndIf
\EndFunction
\end{algorithmic}
\end{algorithm}

\begin{theorem} Suppose that $\OPT \leq \phi$, \Cref{alg:main} outputs a clustering $\mathcal{C}$ with $\obj(\mathcal{C})\leq (3+\eta) \phi$. 
\end{theorem}
\begin{proof} Let $\mathcal{C}^{*}$ be an optimal solution so $\obj(\mathcal{C}^{*}) \leq \phi$. In the next subsections, we show the following:
\begin{enumerate}
\item \label{itm:first} (High-Degree Nodes Clustering). For any $u \in \Vhigh$, $\mathcal{L}_u \cap  \Vhigh = \mathcal{C}^{*}_{u} \cap \Vhigh$. 
\item (No Stealing on Low-Degree Nodes). For each $i$, let $C^{*}_i$ be the cluster in $\mathcal{C}^{*}$ such that $L_i \cap \Vhigh = {C}^{*}_{i} \cap  \Vhigh$. We have $C^{*}_i \cap V_i = C^{*}_i \cap \Vlow$. That is, those low-degree vertices in $C^{*}_i$ are not taken by other clusters $C_j$ for $j<i$.
\item (Low-Degree Nodes Inclusion). For any $i$, $C^{*}_i \cap N(u_i) \subseteq C_i$. 
\item (Closeness). \label{itm:last} For any $i$, $|N[u_i] \Delta C_i| \leq \phi$ and $|C^{*}_i \Delta C_{i}| \leq \phi$.
\end{enumerate}
Once we have shown the above, we can see that $\obj(\mathcal{C}) \leq (3+\eta) \phi$ as follows. Consider a component $C_i$ of $\mathcal{C}$. If $C_i$ is a singleton $\{v\}$ with $\deg(v) \leq (3+\eta)\phi$, then obviously, $\rho_{\mathcal{C}}(v) \leq (3+\eta)\phi$. Otherwise, $C_i$ contains some vertex $x$ with $\deg(x) >  (3+\eta)\phi$. By \Cref{itm:first}, there must exist $C^{*}_i$ such that $C_i \cap \Vhigh = {C}^{*}_{i} \cap \Vhigh$.

Let $v$ be any vertex in $C_i$. We will show that $\rho_{\mathcal{C}}(v) \leq 3\phi$. Suppose that $v \in C^{*}_i \cap C_i$, we have: 
\begin{align*}
\rho_{\mathcal{C}}(v) &= |N[v] \Delta C_i|
\leq |N[v] \Delta C^{*}_i | + |C^{*}_i \Delta C_i|  && \mbox{by \Cref{lem:triangleineq}}\\
&= \rho_{\mathcal{C^{*}}}(v) + |C^{*}_i \Delta C_i| \leq \phi + \phi  = 2\phi 
\end{align*}
Otherwise, if $v \notin C^{*}_i \cap C_i$ then it must be the case that $v \in C_i \setminus C^{*}_i$. In such a case, $v$ must be a vertex in $R(u_i)$ added to $C_i$ in Line \ref{ln:R} to Line \ref{ln:C}. This implies $|N[v] \Delta N[u_i]| \leq_{\eta/2} 2\phi$ and thus $|N[v] \Delta N[u_i]| \leq (1+\eta/2)\cdot2\phi$. Therefore:
\begin{align*}
\rho_{\mathcal{C}}(v) = |N[v] \Delta C_i| 
&\leq |N[v] \Delta N[u_i]| + |N[u_i] \Delta C_i| && \mbox{by \Cref{lem:triangleineq}} \\
&\leq (1+\eta/2)\cdot 2\phi + \phi = (3+\eta)\phi
\end{align*}
\end{proof}

\begin{remark} To get a $(3+\epsilon)$-approximation algorithm, $O(\log n)$ calls of  \Cref{alg:lp} are sufficient, as we can perform a binary search on $\phi$ in the range of $[0,n]$ with $\eta=\epsilon$ and take the solution output by the algorithm with the smallest $\phi$. For a $3$-approximation algorithm, it can also be achieved within $O(\log n)$ calls of \Cref{alg:lp} by setting $\eta = 0$. \end{remark}

We prove each of the four items in the proof in the following subsections. Note that the no-stealing property of low-degree vertices (\Cref{thm:nostealing} in \Cref{sec:nostealing}) is the main structural result, which not only leads to a guarantee on the approximation ratio, but it is also crucial in the design of efficient algorithms in the subsequent sections. 

\subsection{High-Degree Nodes Clustering}
Recall that $\phi$ is our guess of the upper bound of the optimal solution.  In this subsection, we show if $\phi$ is indeed such an upper bound, then there is a unique way to form clustering on nodes with degrees greater than $(3+\eta)\phi$ in the optimal solution. Moreover, in the algorithm, our construction of clustering on those nodes aligns with the way. The following two lemmas are observed by \cite{heidrich20244}.

\begin{lemma}\label{lem:diffcluster} Suppose that $|N[x] \cap N[y]| > t$. If there is a clustering $\mathcal{C}'$ such that $x$ and $y$ are in different clusters, then $\obj(\mathcal{C}') > t/2$. \end{lemma}
\begin{proof}
\begin{align*}
t < |N[x] \cap N[y]| \leq |N[x] \cap N[y] \setminus \mathcal{C'}_{y}| + |N[x] \cap N[y] \setminus \mathcal{C'}_{x}| \leq \rho_{\mathcal{C}'}(y) + \rho_{\mathcal{C}'}(x)
\end{align*}
Therefore, $\obj(\mathcal{C}') \geq \max(\rho_{\mathcal{C'}}(x),  \rho_{\mathcal{C'}}(y)) \geq (\rho_{\mathcal{C'}}(x) + \rho_{\mathcal{C'}}(y) )/2 > t/2$.
\end{proof}

\begin{lemma}\label{lem:samecluster} Suppose that $|N[x] \Delta N[y]| > t$. If there is a clustering $\mathcal{C}'$ such that $x$ and $y$ are in the same clusters, then $\obj(\mathcal{C}') > t/2$. \end{lemma}
\begin{proof}
Let $C$ be the cluster containing $x$ and $y$. By \Cref{lem:triangleineq}, we have: 
\begin{align*}
t &< |N[x] \Delta N[y]| \leq |N[x]\Delta C| + |N[y] \Delta C|= \rho_{\mathcal{C'}}(x) + \rho_{\mathcal{C'}}(y) 
\end{align*}
Therefore, $\obj(\mathcal{C}') \geq \max(\rho_{\mathcal{C'}}(x),  \rho_{\mathcal{C'}}(y)) \geq (\rho_{\mathcal{C'}}(x) + \rho_{\mathcal{C'}}(y) )/2 > t/2$.
\end{proof}

\begin{lemma}\label{lem:highdegreeclustering}
Let $\mathcal{C}^{*}$ be a clustering with $\obj(\mathcal{C}^{*}) \leq \phi$. For any $u \in \Vhigh$, $\mathcal{L}_u \cap \Vhigh = \mathcal{C}^{*}_{u} \cap \Vhigh$. 
\end{lemma}
\begin{proof}
Suppose on the contrary that $\mathcal{L}_u \cap \Vhigh \neq \mathcal{C}^{*}_{u} \cap \Vhigh$. Then there exists some node $x$ such that $x \in (\mathcal{L}_u \cap \Vhigh) \Delta (\mathcal{C}^{*}_{u} \cap \Vhigh)$.

Suppose that there exists a node $x$ such that $x \in \mathcal{L}_u \cap \Vhigh$ but  $x \notin \mathcal{C}^{*}_{u} \cap \Vhigh$. Then there must exist $y \in \mathcal{C}^{*}_{u}$ such that $(x,y) \in E'$. This implies $|N[x] \Delta N[y]| \leq_{\eta} 2\phi$ and so $|N[x] \Delta N[y]| \leq (1+\eta) 2\phi$. Therefore,
\begin{align*}
|N[x] \cap N[y]| &= (\deg(x) +1 + \deg(y)+ 1 - |N[x] \Delta N[y]|)/2 \\
&> ((3+\eta)\phi + 1 + (3+\eta)\phi + 1 - (1+\eta)2\phi ) / 2 \\
&> (6\phi + 2  - 2\phi)/2 = 2\phi + 1
\end{align*}
By \Cref{lem:samecluster}, $\obj(\mathcal{C}^{*}) > \phi$, a contradiction.

Otherwise, it must be the case that $x \in \mathcal{C}^{*}_{u} \cap \Vhigh$ but $x \notin \mathcal{L}_u \cap \Vhigh$. In this case, there exists $y \in \mathcal{C}^{*}_u$ such that $(x,y) \notin E'$. This implies that $|N[x] \Delta N[y]| \nleq_{\eta} 2\phi$, which in turns implies $|N[x] \Delta N[y]| > 2\phi$. By \Cref{lem:samecluster}, $\obj(\mathcal{C}^{*}) > \phi$, a contradiction.

\end{proof}

\subsection{No Stealing on Low-Degree Nodes}\label{sec:nostealing}
In the algorithm, low-degree nodes (i.e.~nodes with degree at most $(3+\eta)\phi$) are added to the clusters formed by high-degree nodes iteratively. In this subsection, we show that if a low-degree node degree node $y$ belongs to $C^{*}_i$ for some $i$, then it will not be included in $C_j$ for $j<i$. This implies that $y \in V_i$, the candidate set of vertices to be added $C_i$. Then in the next subsection, we show that it will be added to $C_i$. 

\begin{theorem}\label{thm:nostealing}
Let $\mathcal{C}^\ast$ be a clustering with $\text{obj}(\mathcal{C}^{*}) \le \phi.$ Let $u$ and $v$ be vertices of degree greater than $(3+\eta)\phi$ where $\mathcal{C}^{*}_u \neq \mathcal{C}^{*}_v$, and $w$ be a vertex with $\text{deg}(w) \leq (3+\eta)\phi$. If $w \in \mathcal{C}^{*}_u$ then $|N[v] \Delta N[w]| > (1+\eta)2\phi$ and so $|N[v] \Delta N[w]| \nleq_{\eta} 2\phi$.
\end{theorem}

\begin{figure}
\centering
\includegraphics[scale=0.45]{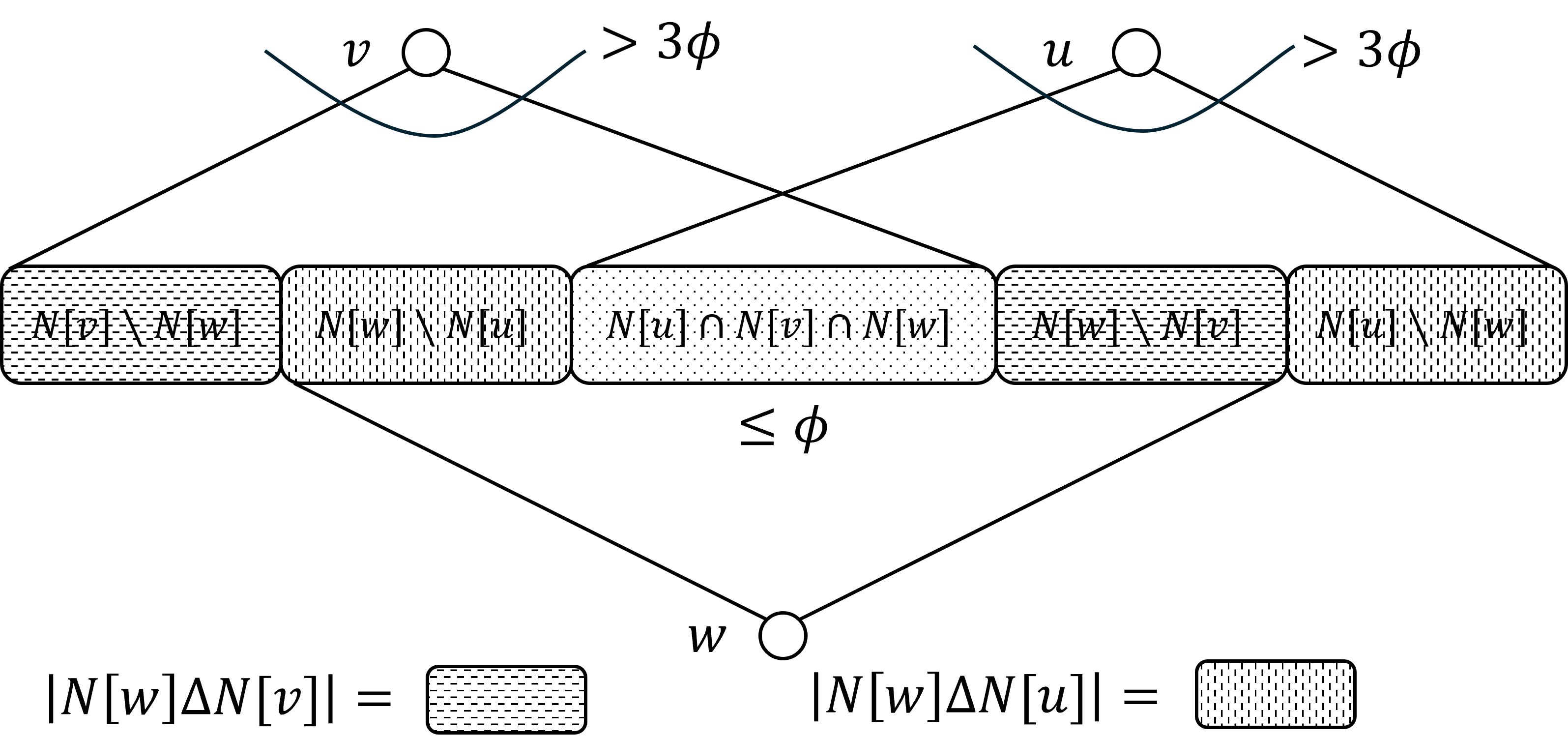}
\caption{A pictorial illustration of the proof of \Cref{lem:nostealingeasy} when $\eta = 0$}\label{fig:1}
\end{figure}

A representative case that illustrates the intuition of why the theorem holds is when $|N[u]\cap N[v] \cap N[w]| \leq \phi$, i.e.~the three sets have small intersections. Since they have small intersection, together with the fact that $u$ and $v$ have degrees greater than $(3+\eta) \phi$, it cannot be the case that both the symmetric differences $N[u] \Delta N[w]$ and $N[v] \Delta N[w]$ are small, as illustrated in \Cref{fig:1}. We will refer the case that $|N[u] \cap N[v] \cap N[w]| \leq \phi$ as the {\it easy case}. 


When the three sets have a large intersection, with a more sophisticated argument on the relations among $N[u], N[v]$, and $N[w]$, it can also be shown that $N[w]$ and $N[u]$ will not intersect a lot. We will refer to the condition that $|N[u] \cap N[v] \cap N[w]| > \phi$ as {\it the hard case}.
\paragraph{The easy case.} We will first show the theorem for the easy case as follows. 
\begin{lemma}\label{lem:nostealingeasy}
Let $\mathcal{C}^\ast$ be a clustering with $\text{obj}(\mathcal{C}^{*}) \le \phi.$ Let $u$ and $v$ be vertices of degree greater than $(3+\eta)\phi$ where $\mathcal{C}^{*}_u \neq \mathcal{C}^{*}_v$, and $w$ be a vertex with $\text{deg}(w) \leq (3+\eta)\phi$. If $w \in \mathcal{C}^{*}_u$ and $|N[v] \cap N[u] \cap N[w]| \leq \phi$, then $|N[v] \Delta N[w]| > (1+\eta)2\phi$ and so $|N[v] \Delta N[w]| \nleq_{\eta} 2\phi$.
\end{lemma}
\begin{proof} We will show that $|N[v] \Delta N[w]| + |N[u] \Delta N[w]|$ is greater than $(1+\eta)4\phi$. \Cref{fig:1} gives a high level illustration of why this should be true. To show this formally, first note that: 
\begin{align*}
&|N[v] \cap N[w] \cap N[u]| + |N[w] \setminus N[u]| + |N[v] \setminus N[w]| \\
&\geq |N[v] \cap N[w] \cap N[u]| + |(N[v] \cap N[w]) \setminus N[u]| + |N[v] \setminus N[w]| \\
&= |N[v] \cap N[w] |+ |N[v] \setminus N[w]| = |N[v]|
\end{align*}
Re-arranging, we have:
\begin{align}
|N[w] \setminus N[u]| + |N[v] \setminus N[w]| &\geq |N[v]| - |N[v] \cap N[w] \cap N[u]| > (3+\eta)\phi - \phi = (2+\eta)\phi \label{eqn:dif1}
\end{align}
By the same reasoning, we have
\begin{align}
|N[w] \setminus N[v]| + |N[u] \setminus N[w]| > (2+\eta)\phi \label{eqn:dif2}
\end{align}
Now consider the following:
\begin{align*}
&|N[v] \Delta N[w]| + |N[u] \Delta N[w]| \\
&= |N[v] \setminus N[w]| + |N[w] \setminus N[v]| + |N[u] \setminus N[w]| + |N[w] \setminus N[u]| \\
&> (2+\eta)\phi + (2+\eta)\phi && \mbox{by (\ref{eqn:dif1}) and (\ref{eqn:dif2})} \\
&= (4+2\eta)\phi
\end{align*}
Therefore, assume to the contrary that if $|N[v] \Delta N[w]| \leq_{\eta} 2\phi$ and so $|N[v] \Delta N[w]| \leq (1+\eta)2\phi$ then it must be the case that $|N[u] \Delta N[w]| > (4+2\eta)\phi - |N[v] \Delta N[w]| \geq 2\phi$. By the fact that $w \in \mathcal{C}^{*}_v$ and \Cref{lem:samecluster}, $\obj(\mathcal{C}^{*}) > \phi$, a contradiction.
\end{proof}

\paragraph{The hard case} Now we consider the case where $|N[w] \cap N[u] \cap N[v]| > \phi$. To illustrate the idea of proof, suppose that $|N[w] \cap N[u] \cap N[v]| = \phi + t$ for some $t>0$. If we follow the same argument as \Cref{lem:nostealingeasy}, we would only be able to show that $|N[w] \Delta N[v]| + |N[w] \Delta N[u]| > (4+2\eta)\phi - 2t$ as opposed to $(4+2\eta)\phi$. 

However, if this is the case, we can show that $|N[w] \Delta N[u]| \leq 2(\phi-t)$, as stated in \Cref{lem:largeintersection}. This would make $|N[w] \Delta N[v]| > (1+\eta)2\phi$. 

\begin{figure}
\centering
\includegraphics[scale=0.45]{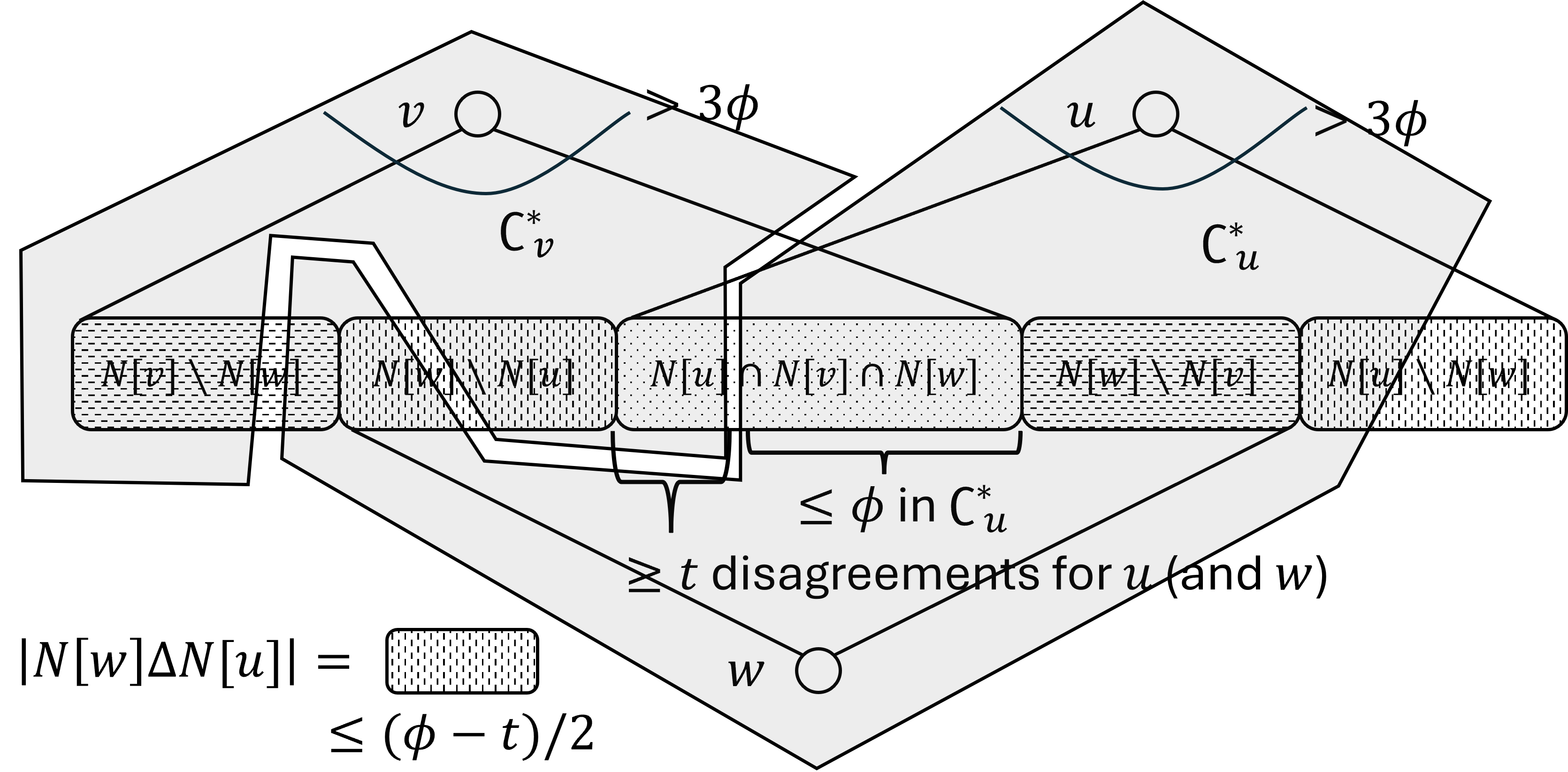}
\caption{A pictorial illustration of the proof of \Cref{lem:largeintersection} when $\eta = 0$}\label{fig:2}
\end{figure}

The high-level idea of the proof of \Cref{lem:largeintersection} is illustrated in \Cref{fig:2}. First, we observe that at most $\phi$ vertices in $|N[u] \cap N[v] \cap N[w]|$ can be contained in $\mathcal{C}^{*}_u$; otherwise the vertex $v$ would have disagreements greater than $\phi$. This implies the contribution of disagreement from $N[u] \cap N[w]$ to $u$ (and to $w$) is already at least $t$. If the symmetric difference $|N[u] \Delta N[w]|$ is too large, i.e.~larger than $2(\phi -t)$, then it would force the disagreements of either $u$ or $w$ to be too large.

\begin{lemma}\label{lem:largeintersection}
Let $\mathcal{C}^\ast$ be a clustering with $\text{obj}(\mathcal{C}^{*}) \le \phi.$ Let $u$ and $v$ be vertices of degree greater than $(3+\eta)\phi$ where $\mathcal{C}^{*}_u \neq \mathcal{C}^{*}_v$, and $w \in \mathcal{C}^{*}_u$ be a vertex with $\text{deg}(w) \leq (3+\eta)\phi$. Suppose that $|N[w] \cap N[v] \cap N[u]| = \phi + t$ for some $t>0$. Then, $|N[w] \Delta N[u]| \le 2(\phi - t)$.
\end{lemma}
\begin{proof}
Let $S = N[w] \cap N[v] \cap N[u]$. First we claim that at most $\phi$ vertices in $S$ can be in $\mathcal{C}^{*}_u$. This in turns implies that vertices in $S$ contributes at least $t$ disagreements to vertex $u$ and vertex $w$. To see why this holds, suppose to the contrary that $|S \cap \mathcal{C}^{*}_u| > \phi$. Since $S \subseteq N[v]$ and $\mathcal{C}^{*}_v \neq \mathcal{C}^{*}_u$, it must be the case that $\rho_{\mathcal{C}^{*}}(v) \geq |N[v] \setminus C^{*}_v| \geq |S \setminus C^{*}_v| \geq |S \cap \mathcal{C}^{*}_u| > \phi$, a contradiction. Therefore, we conclude that:
\begin{align}|S \setminus \mathcal{C}^{*}_u| \geq t\end{align}

Now note that the disagreements of $u$ and $w$ are at least:
\begin{align*}
\rho_{\mathcal{C}^{*}}(u) &\geq |S \setminus \mathcal{C}^{*}_u| + |(N[w] \setminus N[u]) \cap \mathcal{C}^{*}_u| + |(N[u] \setminus N[w]) \setminus \mathcal{C}^{*}_u| \\
\rho_{\mathcal{C}^{*}}(w) &\geq |S \setminus \mathcal{C}^{*}_u| + |(N[u] \setminus N[w]) \cap \mathcal{C}^{*}_u| + |(N[w] \setminus N[u]) \setminus \mathcal{C}^{*}_u|
\end{align*}

Using the fact $\rho_{\mathcal{C}^{*}}(u), \rho_{\mathcal{C}^{*}}(w) \leq \phi$ and adding the above two inequalities together, we have:
\begin{align*}
2\phi &\geq \rho_{\mathcal{C}^{*}}(u) + \rho_{\mathcal{C}^{*}}(w) \\
&\geq 2|S \setminus \mathcal{C}^{*}_u| + |(N[w] \setminus N[u]) \cap \mathcal{C}^{*}_u| + |(N[u] \setminus N[w]) \setminus \mathcal{C}^{*}_u| + \\
& \hspace{25mm} |(N[u] \setminus N[w]) \cap \mathcal{C}^{*}_u| + |(N[w] \setminus N[u]) \setminus \mathcal{C}^{*}_u| \\
&= 2|S \setminus \mathcal{C}^{*}_u| + |(N[w] \Delta N[u]) \cap \mathcal{C}^{*}_u | + |(N[u] \Delta N[w]) \setminus \mathcal{C}^{*}_u| \\
&= 2|S \setminus \mathcal{C}^{*}_u| + |(N[w] \Delta N[u])| \\
&\geq 2t + |(N[w] \Delta N[u])|
\end{align*}

Therefore, $|N[w] \Delta N[v]|\leq 2(\phi - t)$.
\end{proof}

\begin{lemma}\label{lem:nostealinghard}
Let $\mathcal{C}^\ast$ be a clustering with $\text{obj}(\mathcal{C}^{*}) \le \phi.$ Let $u$ and $v$ be vertices of degree greater than $(3+\eta)\phi$ where $\mathcal{C}^{*}_u \neq \mathcal{C}^{*}_v$, and $w$ be a vertex with $\text{deg}(w) \leq (3+\eta)\phi$. If $w \in \mathcal{C}^{*}_u$ and $|N[v] \cap N[u] \cap N[w]| > \phi$, then $|N[v] \Delta N[w]| > (1+\eta)2\phi$ and so $|N[v] \Delta N[w]| \nleq_{\eta} 2\phi$.
\end{lemma}

\begin{proof}
By (\ref{eqn:dif1}) in the proof \Cref{lem:nostealingeasy}, we have:
\begin{align*}
|N[w] \setminus N[u]| + |N[v] \setminus N[w]| &\geq |N[v]| - |N[v] \cap N[w] \cap N[u]|  \\
& > (3+\eta) \phi - (\phi + t) = (2+\eta)\phi - t \\
|N[w] \setminus N[v]| + |N[u] \setminus N[w]| &\geq |N[u]| - |N[v] \cap N[w] \cap N[u]|  \\
&> (3+\eta) \phi - (\phi + t) = (2+\eta)\phi - t
\end{align*}
Therefore, 
\begin{align*}
|N[v] \Delta N[w]| + |N[u] \Delta N[w]| &= |N[v] \backslash N[w]| + |N[w] \backslash N[v]| + |N[u] \backslash N[w]| + |N[u] \backslash N[w]| \\
&> (4+2\eta)\phi - 2t
\end{align*}
By \Cref{lem:largeintersection}, we have $|N[u] \Delta N[w]| \geq 2(\phi - t) $, therefore:
$$|N[v] \Delta N[w]| > (4+2\eta)\phi - 2t - 2(\phi - t) = (1+\eta)2\phi$$
Hence, $|N[v] \Delta N[w]| \nleq_{\eta} 2\phi$.
\end{proof}
\Cref{lem:nostealingeasy} and \Cref{lem:nostealinghard} complete the proof of \Cref{thm:nostealing}. \Cref{thm:nostealing} immediately implies the following:
\begin{corollary}\label{cor:nostealing}
Let $C^{*}_i$ be the cluster in $\mathcal{C}^{*}$ such that $L_i \cap \Vhigh = {C}^{*}_{i} \cap  \Vhigh$. For any $i$, $C^{*}_i \cap V_i = C^{*}_i \cap \Vlow$.
\end{corollary}
\subsection{Low-Degree Nodes Inclusion}
Here, we show that all the vertices in $\mathcal{C}^{*}_i$ have similar neighborhoods with $u_i$. As a result, if they are neighbors of $u_i$, they will be added to $C_i$. 
\begin{lemma}\label{lem:inclusion} Let $C^{*}_i$ be the cluster in $\mathcal{C}^{*}$ such that $L_i \cap \Vhigh = {C}^{*}_{i} \cap \Vhigh$. For any $i$, $C^{*}_i \cap N(u_i) \subseteq C_i$. 
\end{lemma}

\begin{proof}
If $w \in C^{*}_i \cap \Vhigh$, then $w \in C_i$ because $C_i = L_i \cup R(u_i)$ and $L_i \cap \Vhigh = C^{*}_i \cap \Vhigh$ by assumption. Now it remains to consider the case $w \in C^{*}_i \cap \Vlow \cap N(u_i)$. It suffices for us to show that $w \in R(u_i)$, where $R(u_i) = \{ w \in V_i \cap N(u_i) \mid |N[w] \Delta N[u_i] | \leq_{\eta} 2\phi  \}$, as $R(u_i)$ will be added to $C_i$.

By \Cref{cor:nostealing}, we have $w \in C^{*}_i \cap V_i \cap N(u_i)$, which means $w$ is not going to be stolen by other clusters. To show that $w \in C_i$, now it suffices to show that $|N[w] \Delta N[u_i]| \leq_{\eta} 2\phi$ always holds. This is indeed true, because:
\begin{align*}
|N[w] \Delta N[u_i]| &\leq |N[w] \Delta C^{*}_i| + |C^{*}_i \Delta N[u_i]| && \mbox{by \Cref{lem:triangleineq}} \\
&\leq \phi + \phi && \mbox{$u_i, w \in C^{*}_i,\rho_{\mathcal{C}^{*}}(w), \rho_{\mathcal{C}^{*}}(u_i) \leq \phi$} \\
&\leq 2\phi
\end{align*}
\end{proof}
\subsection{Closeness}
In the following, we will show that the cluster we constructed $C_i$ will be similar to $C^{*}_i$ and $N[u_i]$. Intuitively, this holds because the low-degree part of $C_i$ will be sandwiched between the low-degree part of $C^{*}_i \cap N[u_i]$ and $N[u_i]$, i.e.~$C^{*}_i \cap N[u_i] \cap \Vlow \subseteq C_i \cap \Vlow \subseteq N[u_i] \cap \Vlow$. Also, we know that $N[u_i]$ and $C^{*}_i$ are close (i.e.~$|N[u_i] \Delta C_i| \leq \phi$), so somehow $C_i$ cannot be too far away from $C^{*}_i$ and $N[u_i]$. Note that the high-degree parts of $C_i$ and $C^{*}_i$ coincide. 
\begin{restatable}{lemma}{closeness}\label{lem:closeness} Let $C^{*}_i$ be the cluster in $\mathcal{C}^{*}$ such that $L_i \cap \Vhigh = {C}^{*}_{i} \cap \Vhigh$. For any $i$, $|N[u_i] \Delta C_i| \leq \phi$ and $|C^{*}_i \Delta C_{i}| \leq \phi$. 
\end{restatable}

\begin{proof}
We first show that $|N[u_i] \Delta C_i| \leq \phi$.
\begin{align}
|\Vlow \cap (N[u_i] \Delta C_i)| &= |\Vlow \cap (N[u_i] \setminus C_i)| + |\Vlow \cap (C_i \setminus N[u_i])| \nonumber \\
&= |\Vlow \cap (N[u_i] \setminus C_i)|  && (\Vlow \cap C_i) \subseteq N[u_i]  \nonumber  \\
&\leq |\Vlow \cap (N[u_i] \setminus C^{*}_i)|  && \mbox{by \Cref{lem:inclusion}}   \nonumber \\
&\leq |\Vlow \cap (N[u_i] \Delta C^{*}_i)|  \label{eqn:diffneighborCstar1}
\end{align}

Moreover, since $C^{*}_i \cap \Vhigh = C_i \cap \Vhigh$, it must be the case that 
\begin{align}
|(N[u_i] \Delta C_i) \cap \Vhigh| = |(N[u_i] \Delta C^{*}_i) \cap \Vhigh|  \label{eqn:diffneighborCstar2}
\end{align}

Therefore,
\begin{align*}
|N[u_i] \Delta C_i| &= |(N[u_i] \Delta C_i) \cap \Vlow| + |(N[u_i] \Delta C_i) \cap \Vhigh| \\
&\leq |(N[u_i] \Delta C^{*}_i) \cap \Vlow| + |(N[u_i] \Delta C^{*}_i) \cap \Vhigh| && \mbox{by (\ref{eqn:diffneighborCstar1}) and (\ref{eqn:diffneighborCstar2})} \\
& = |N[u_i] \Delta C^{*}_i| \leq \phi  && \mbox{$u_i \in C^{*}_i$ and $\rho_{C^{*}}(u_i) \leq \phi$}
\end{align*}

Now we show that $|C^{*}_i \Delta C_{i}| \leq \phi$. 
\begin{align*}
&|C^{*}_i \Delta C_{i}| \\
&= |\Vlow \cap (C^{*}_i \Delta C_{i})| && C^{*}_i \cap \Vhigh = C_i \cap \Vhigh \\
&= |\Vlow \cap (C^{*}_i \setminus C_{i})| + |\Vlow \cap (C_i \setminus C^{*}_{i})| \\
&= |\Vlow \cap (C^{*}_i\cap N[u_i] \setminus C_{i})| + |\Vlow \cap (C^{*}_i \setminus N[u_i] \setminus C_{i})| \\
& \hspace{50mm} + |\Vlow \cap (C_i \setminus C^{*}_{i})| \\
&= |\Vlow \cap (C^{*}_i\cap N[u_i] \setminus C_{i})| + |\Vlow \cap (C^{*}_i \setminus N[u_i])| \\
& \hspace{50mm} + |\Vlow \cap (C_i \setminus C^{*}_{i})| && \Vlow \cap C_i \subseteq \Vlow \cap N[u_i]\\
&= |\Vlow \cap (C^{*}_i \setminus N[u_i])| + |\Vlow \cap (C_i \setminus C^{*}_{i})| && C^{*}_i\cap N[u_i] \subseteq C_i \\
&\leq |\Vlow \cap (C^{*}_i \setminus N[u_i])| + |\Vlow \cap (N[u_i] \setminus C^{*}_{i})| && \Vlow \cap C_i \subseteq \Vlow \cap N[u_i] \\
&= |\Vlow \cap (C^{*}_i \Delta N[u_i])| \leq |C^{*}_i \Delta N[u_i]| \leq \phi
\end{align*}

\end{proof}

\section{Efficient Algorithms for the Sequential and MPC Models} \label{sec:efficientimplementation}
To make the algorithm efficient in the sequential and MPC settings, we need to address several challenges. First, we need to compute the graph $E'$ for high-degree nodes. Second, we need to be able to conduct approximate neighborhood similarity queries efficiently. We will address these issues one by one.

\subsection{Computing $\mathcal{L}$ for High-Degree Nodes}

In this section, we show that although $E'$ may be significantly larger than $E^{+}$, it is sufficient to make only $O(|E^{+}|) = O(m)$ neighborhood similarity queries to identify the high-degree clusters. \Cref{alg:highdegclustering} describes how to form a clustering $\mathcal{F}$ for high-degree vertices by conducting $O(m)$ similarity queries. 

\paragraph{Algorithm Description} In \Cref{alg:highdegclustering}, we first perform $O(m)$ neighborhood queries to construct $\Gsim = (V, \Esim)$, where $\Esim \subseteq E^{+}$ consists of those endpoints who have similar neighborhoods. Then we assign every vertex $u$ with a unique identifier, $\ID(u)$. By sending the ID of high-degree every vertex to its neighbors in $\Gsim$, every vertex $u$ can learn ID of the high-degree vertex with the smallest ID in its closed neighborhood in $\Gsim$ (Line \ref{ln:learnID}) and then set $\min(u)$ to be such an ID. Now, every vertex $u$ (including low-degree vertices) sends a token with value $\min(u)$ to all the neighbors of $u$ in $\Gsim$. Then, if a vertex $u$ recieved at least $\phi + 1$ tokens of the same value, it will set its cluster ID, $cluster(u)$, to be the minimum value that occurs at least $\phi + 1$ times. Finally, we assign all vertices with the same cluster ID will be in the same cluster. 

In \Cref{lem:efficienthighdeg}, we show that if $\OPT \leq \phi$, then the clustering $\mathcal{F}$ constructed is exactly the same as $\mathcal{L}$. The proof is based on showing that if two high-degree vertices $u,v$ are in the same cluster in $\mathcal{C}^{*}$, then there will be at least $\phi + 1$ disjoint paths of length two in $\Gsim$. 


\begin{lemma}\label{lem:efficienthighdeg} Suppose that $\OPT \leq \phi$. \Cref{alg:highdegclustering} outputs a clustering $\mathcal{F}$ such that $\mathcal{F} = \mathcal{L}$. \end{lemma}

\begin{proof}
Let $\mathcal{C}^{*}$ be a clustering with $\obj(\mathcal{C}^{*})\leq \phi$.  Let $u$ be a vertex in $\Vhigh$. 
Let $u_{*}$ be the vertex with the minimum ID in $\mathcal{L}_u$. We will show that 
\begin{enumerate}
    \item Every vertex $x \in \mathcal{L}_u \cap \Vhigh$ will receive at least $\phi + 1$ tokens of values $\ID(u_{*})$ in Line \ref{ln:receivemany}.
    \item In addition, if $x$ received at least $\phi + 1$ tokens of value $T$, then $T \geq \ID(u_{*})$. 
\end{enumerate} 

Once these two points are established, every vertex $x \in \mathcal{L}_u \cap \Vhigh$ will be assigned $\texttt{cluster}(x) = \ID(u_*)$. Consequently, this ensures that $\mathcal{F}_u = \mathcal{L}_u = \mathcal{C}^*_u \cap \Vhigh$.

\begin{algorithm}[ht!]
\caption{{\textsc{HighDegreeClustering}}$(G^+=(V, E^+), \phi, \eta)$\label{alg:highdegreecluster}\\
\textbf{Input}: A graph $G^{+}$, a parameter $\phi$, a parameter $0 \leq \eta < 1$. \\
\textbf{Output}: A clustering $\mathcal{F} = \{L_i\}_{i=1}^{|\mathcal{L}|}$ of $\Vhigh$ such that $\mathcal{F} = \mathcal{L}$. }
\label{alg:highdegclustering}
\begin{algorithmic}[1]
\Function {\textsc{HighDegreeClustering}}{$G^+=(V, E^+), \phi, \eta$} 
    \State Let $\Gsim = (V,\Esim)$, where $\Esim = \{uv \in E^{+} \mid |N[u] \Delta N[v] | \leq_{\eta} 2\phi \}$. \label{ln:2} 
    \State Assign each vertex $V$ with an unique ID, $\ID(v)$.

    \State \label{ln:learnID}Every vertex $v \in V$ sets $\min(v) \leftarrow \min_{x \in N_{\Gsim}[v] \cap \Vhigh}\ \ID(x) $
    \State Every vertex $v \in V$ sends a token of value $\min(v)$ to the neighbor of $v$ in $\Gsim$.
    \State \label{ln:receivemany}If a vertex $v \in \Vhigh$ receives at least $\phi + 1$ tokens of the same value, set $cluster(v)$ to be the minimum such value. 
    
    \State Let $\mathcal{F}$ be a clustering where each cluster is formed by vertices with the same $cluster$ values. 
\EndFunction
\end{algorithmic}
\end{algorithm}

Now we start with the first point, since $x, u_{*} \in \mathcal{C}^{*}_u$, $|N[u_{*}] \Delta N[x]| \leq 2\phi$; for otherwise $\obj(\mathcal{C}^{*}) > \phi$ by \Cref{lem:samecluster}. This implies:
\begin{align*} 
|N[u_{*}] \cap N[x]| &=  (|N[x]| + |N[u_{*}]| - |N[u_{*}] \Delta N[x]|)/2 \\
&\geq (3\phi + 2  + 3\phi + 2 - 2\phi )/2 = 2\phi + 2
\end{align*}
Also, since $u_{*} \in \mathcal{C}^{*}_{u}$ and by definition of $\rho_{\mathcal{C}^{*}}(u_{*})$, we have: 
\begin{align}
\phi &\geq \rho_{\mathcal{C}^{*}}(u_{*}) \geq |N[u_{*}]| - |(\mathcal{C}^{*}_{u} \cap N[u_{*}])|\label{eqn:diff}
\end{align}
Therefore, 
\begin{align*}
|(\mathcal{C}^{*}_{u} \cap N[u_{*}]) \cap N[x]| &= |N[x] \cap N[u_{*}]| - |N[x] \cap (N[u_{*}] \setminus (\mathcal{C}^{*}_{u} \cap N[u_{*}]))| \\
&\geq |N[x] \cap N[u_{*}]| - |N[u_{*}] \setminus (\mathcal{C}^{*}_{u} \cap N[u_{*}])| \\
&\geq (2\phi + 2) - \phi && \mbox{by (\ref{eqn:diff})}\\
&\geq \phi + 2
\end{align*}
Hence:
$$|(\mathcal{C}^{*}_{u} \cap N[u_{*}]) \cap N(x)| \geq |(\mathcal{C}^{*}_{u} \cap N[u_{*}]) \cap N[x]| - 1 \geq \phi + 1$$
Note that all vertices in $(\mathcal{C}^{*}_{u} \cap N[u_{*}])$ must have their $\min(\cdot)$ values equal to $\ID(u_{*})$. This is because if $v \in (\mathcal{C}^{*}_{u} \cap N[u_{*}])$ and $v \in \Vhigh$, then $(N_{\Gsim}(v) \cap \Vhigh) \subseteq \mathcal{C}^{*}_{u}$.  If $v \in (\mathcal{C}^{*}_{u} \cap N[u_{*}])$ and $v \notin \Vhigh$, then it must be the case that $v \in N_{\Gsim}(u_{*})$. Moreover, $N_{\Gsim}[v] \cap \Vhigh$ cannot contain any cluster nodes outside $\mathcal{C}^{*}_{u}$ by \Cref{thm:nostealing}. Thus, $\min(v) = \ID(u_{*})$.  This implies $x$ will receive at least $\phi + 1$ tokens with value $\ID(u_{*})$.

Next we show that if there is a value $T$ such that $x$ receives from at least $\phi+1$ tokens, then $T \geq \ID(u_{*})$. Suppose to the contrary that $T < \ID(u_{*})$.  First note that if $v \in N(x)$ has $\min(v) = T$, then it cannot be the case that $v \in \Vhigh$. Otherwise, $u_{*}$, $v$, and a vertex whose ID is $T$,  would all be in $\mathcal{C}^{*}_{u}$, which contradict with the fact $\ID(u_{*})$ is the smallest ID in $\mathcal{L}_u = \mathcal{C}^{*}_{u} \cap \Vhigh$. Therefore, it must be the case that vertices $v \in N(x)$ with $\min(v) = T$ are all in $\Vlow$.  

Now note that if $v \in \mathcal{C}^{*}_u$ then by \Cref{thm:nostealing}, $v$ can only be connected to vertices in $\mathcal{C}^{*}_u$ in $\Gsim$. In this case, $\min(v) \geq \ID(u_{*})$ so $\min(v) \neq T$ by the assumption of $\ID(u_{*})$. So the only possibility for $v$ to have $\min(v) = T$ is when $v \notin \mathcal{C}^{*}_u$. However, as $\rho_{\mathcal{C}^{*}}(x) \leq \phi$, there are at most $\phi$ neighbors of $x$ not in $\mathcal{C}^{*}_{x}$. This implies $x$ will receive less than $\phi + 1$ tokens whose value equals to $T$.





\end{proof}

\subsection{Approximate Neighborhood Similarity Testing by Random Projection}
In this section, we show that for $0 < \eta < 1$, w.h.p.~the queries $\Delta_{\eta}(x,y,2\phi)$ all $xy \in E^{+}$ can be answered in $O(m\log n / \eta^2)$ time and thus $\Gsim$ can be constructed in $O(m\log n / \eta^2)$ time.

\begin{definition} Given a vertex $x$, let $\vec{N}[x] \subseteq \{0,1\}^{n}$ denote the characteristic vector of $N[x]$. \end{definition}
\begin{lemma} Given $\epsilon > 0$, let $k = C \cdot (\log n / \epsilon^2)$ for some large enough constant $C>0$. Let $A$ be a $k \times n$ matrix where each entry is drawn from $\{-1, +1\}$ uniformly at random. W.h.p.~for every two vertices $x$ and $y$, we have:
$$ (1-\epsilon) \cdot |N[x] \Delta N[y] | \leq ||A \cdot \vec{N}[x] - A \cdot \vec{N}[y]||^2_{2}/k \leq (1+\epsilon) \cdot  |N[x] \Delta N[y] |$$
\end{lemma}
\begin{proof} By the Johnson-{L}indenstrauss lemma \cite{Achlioptas03,JL84}, w.h.p.~for every $x$ and $y$, we have: 
$$(1-\epsilon) \cdot ||\vec{N}[x] - \vec{N}[y]||^2_{2} \leq ||A \cdot \vec{N}[x] - A \cdot \vec{N}[y]||^2_{2}/k \leq (1+\epsilon) \cdot ||\vec{N}[x] - \vec{N}[y]||^2_{2}$$

The lemma follows by observing that $|N[x] \Delta N[y]| = ||\vec{N}[x] - \vec{N}[y]||^2_{2}$. 
\end{proof}

\begin{lemma}\label{lem:test}
Let $\epsilon = \Theta(\eta)$ be such that $(1+\eta) = (1+\epsilon)/(1-\epsilon)$, $k = O(\log n /\epsilon^2)$,  and set:
$$\Delta_{\eta}(x,y,t) = \begin{cases} 0 & \mbox{if $||A \cdot \vec{N}[x] - A \cdot \vec{N}[y]||^2_{2}/((1+\epsilon)k) > t$} \\
1 & \mbox{otherwise} \end{cases}$$
W.h.p.~the above implementation returns a correct answer for $\Delta_{\eta}(x,y,t)$.
\end{lemma}
\begin{proof}
It suffices to show that if $|N[x] \Delta N[y]| \leq t$ then w.h.p.~we will set $\Delta_{\eta}(x,y,t)$ to be $0$ and if $|N[x] \Delta N[y]| > (1+\eta)t$, then w.h.p.~we will set $\Delta_{\eta}(x,y,t)$ to be 1.

If $|N[x] \Delta N[y]| \leq t$, then w.h.p. $$||A \cdot \vec{N}[x] - A \cdot \vec{N}[y]||^2_{2}/((1+\epsilon)k) \leq |N[x] \Delta N[y]| \leq t$$
Thus, $\Delta_{\eta}(x,y,t)$ returns 1. 

On the other hand, if $|N[x] \Delta N[y]| > (1+\eta)t$ then w.h.p.
$$||A \cdot \vec{N}[x] - A \cdot \vec{N}[y]||^2_{2}/((1+\epsilon)k) \geq \frac{1-\epsilon}{1+\epsilon}\cdot (|N[x] \Delta N[y]|) > \frac{1-\epsilon}{1+\epsilon}(1+\eta) \cdot t = t$$
Thus, $\Delta_{\eta}(x,y,t)$ returns 0. 
\end{proof}


We have presented the key ingredients that lead to efficient sequential and MPC algorithms. The details of their implementations are presented in \Cref{sec:efficientsequential} and \Cref{sec:efficientmpc}. For the streaming algorithm, different challenges arise. The necessary modifications and details are presented in \Cref{sec:streaming}. 

\subsection{Efficient Sequential Implementation}\label{sec:efficientsequential}
\begin{lemma}\label{lem:efficient_test} For any $0 < \eta < 1$, all neighborhood similarity tests $|N[x] \Delta N[y]| \leq_{\eta} 2\phi$ for $xy \in E^{+}$ can be performed and with correct answers w.h.p.~in $O(\eta^{-2} m \log n)$ time. Therefore, $\Gsim$ can be constructed correctly w.h.p.~in $O(\eta^{-2} m \log n)$ time.\end{lemma}

\begin{proof}
By \Cref{lem:test}, it suffices to show that $||A \cdot \vec{N}[x] - A \cdot \vec{N}[y]||^2_{2}/k$ can be computed for all $xy \in E^{+}$ in $O(\eta^{-2} m \log n)$ time for $k = O(\log n /\eta^2)$, where $A$ is a $k \times n$ matrix drawn from $\{-1, +1\}^{k \times n}$ uniformly at random. 

For each node $v \in V$, we will sample a vector $A_v \in \{-1, +1\}^k$ and so $A = (A_{v_1} \cdots A_{v_n})$. Note that for each $x$, $A \cdot \vec{N}[x] = \sum_{v \in N[x]} A_v$ can be computed in $O(\deg(x)\cdot k)$ time and $A \cdot \vec{N}[x]$ is a $k$-dimension vector. For every edge $xy \in E^{+}$, $A\cdot \vec{N}[x] - A\cdot \vec{N}[y]$ can be computed in $O(k) = O(\eta^{-2} \log n)$ time. Therefore, $\Gsim$ can be constructed in $O(\eta^{-2 }m \log n)$ time.
\end{proof}

\begin{lemma}[Item \ref{itm:main1} of \Cref{thm:thmmain}]
\label{lemma:mainresultsequential}
Given a min-max correlation clustering instance $G = (V, E^{+})$, for any constant $\epsilon > 0$, there exists a randomized sequential algorithm that outputs a clustering that is a $(3 + \epsilon)$-approximation for min-max correlation clustering. This algorithm succeeds with high probability and runs in $O(m \log^2 n / \epsilon^2)$ time.
\end{lemma}
\begin{proof}
By \Cref{lem:efficienthighdeg}, Line \ref{ln:highdegstart}--Line \ref{ln:highdegend} of Algorithm~\ref{alg:lp} can be replaced by \Cref{alg:highdegclustering}, which computes a clustering $\mathcal{F} = \mathcal{L}$ for high-degree graphs provided $\OPT \leq \phi$. Let $\eta = \epsilon$, by \Cref{lem:efficient_test}, $\Gsim$ can be constructed in $O(m \log n /\epsilon^2)$ time. The rest of \Cref{alg:highdegclustering} can be performed in $O(m)$ time sequentially as they involve sending $O(1)$ messages along each edge of $\Gsim$. 

Line \ref{ln:upper}--Line \ref{ln:lower} of Algorithm~\ref{alg:lp} involves $O(m)$ neighborhood similarity tests in total, which can again, be done in $O(m \log n /\epsilon^2)$ time by \Cref{lem:efficient_test}. Line \ref{ln:checking} of Algorithm ~\ref{alg:lp} needs to compute the objective function of the clustering constructed. This can be done in $O(m)$ time by scanning through $N(u)$ for each $u \in G$.
Finally, the binary search on $\phi$ introduces at most $O(\log n)$ calls of \Cref{alg:lp}, so the total running time is $O(m \log^2 n / \epsilon^2)$.
\end{proof}

\exactthreeapprox*
\begin{proof}
Here we implement the algorithm with $\eta = 0$. The arguments are exactly the same with the proof of \Cref{lemma:mainresultsequential}, except that we will perform each neighborhood similarity query (with $\eta = 0$) in $O(D \log D)$ time. A query can be performed in such time, if we pre-build a balanced binary search tree that stores the neighborhood of each vertex, which takes $O(\sum_{u} d(u) \log d(u)) = O(m D \log D)$ time. Since there are at most $O(\log n)$ calls of \Cref{alg:lp} in the binary search of $\phi$ and we perform at most $O(m)$ such queries for each call, the total time is $O(m (D \log D)(\log n))$.
\end{proof}

\subsection{Efficient MPC Implementation}\label{sec:efficientmpc}
In this section, we aim to present the MPC component of Theorem~\ref{thm:thmmain}. More precisely, we aim to show the following lemma,

\begin{lemma}[Item \ref{itm:main2} of \Cref{thm:thmmain}]
\label{lemma:mainresultmpc}
Given a min-max correlation clustering instance $G = (V, E)$, for any constant $\delta > 0, \epsilon$, there exists a randomized MPC algorithm that, in $O(\log(1/\epsilon))$ rounds, outputs a clustering that is a $(3 + \epsilon)$-approximation for min-max correlation clustering. This algorithm succeeds with high probability and uses $O(n^{\delta})$ memory per machine and a total memory of $O(m \log n / \epsilon^2)$.
\end{lemma}

Algorithm~\ref{alg:lp} consists of two parts: constructing clusters $\mathcal{F}$ for $\Vhigh$ (handled by Algorithm~\ref{alg:highdegclustering}) and processing each cluster in $\mathcal{F}$ to manage the low-degree nodes. For Algorithm~\ref{alg:highdegclustering}, we first sample each node with a vector $A_v$, append $A_u$ and $A_v$ to each edge $uv$, and sort all edges by their endpoints. This allows us to compute $A \cdot \vec{N}[x]$ in $O(1 /\delta)$ rounds, as sorting in the MPC model takes $O(1/\delta)$ rounds. In Lines 4--5, each node propagates its information to its neighbors, which can also be done using sorting in $O(1/\delta)$ rounds. Lines 6--7 involve assigning labels, which similarly requires $O(1/\delta)$ rounds.

Once $\mathcal{L} = \mathcal{F}$ is computed, we process all low-degree nodes in Algorithm~\ref{alg:lp}. For each $L_i \in \mathcal{L}$, we randomly choose a node $u_i$. For each edge $wu_i$, we use the $A \cdot \vec{N}[x]$ value to check whether $w \in R(u_i)$. Based on Theorem~\ref{thm:nostealing}, there will be exactly one $u_i$ that adds $w$ to its cluster. Propagating information from $u_i$ to $w$ can be done using sorting, taking $O(1/\delta)$ rounds. 

At the end of Algorithm~\ref{alg:lp}, we compute $\obj(\mathcal{C})$ by appending the cluster information to each node and edge. In summary, Algorithm~\ref{alg:lp} completes in $O(1/\delta)$ rounds and uses $O(m \log n)$ total memory.

A remaining question is how Algorithm~\ref{alg:lp} determines $\phi$, which is provided as input. We can use binary search to estimate $\phi$ close to $\OPT$. While binary search could take $O(\log n)$ rounds with a poor initial guess, Cao, Li, and Ye~\cite{cao2024simultaneously} provided an $O(1)$-round MPC algorithm achieving an $O(1)$-approximation for min-max correlation clustering. Their main result is as follows:

\begin{theorem}[Restatement of Theorem~1.5~in~\cite{cao2024simultaneously}]
There exists an MPC algorithm in the strictly sublinear regime that, given a correlation clustering instance $G = (V, E)$, outputs a clustering that is a $360$-approximation for min-max correlation clustering in $O(1)$ rounds. This algorithm succeeds with high probability and uses a total memory of $O(m \log n)$.
\end{theorem}

We use their algorithm to compute an initial solution $\phi_{s} \leq 360 \OPT$, then maintain an upper bound $\phi_{u} = \phi_{s}$ and a lower bound $\phi_{l} = \phi_{s} / 360 \geq \OPT$ for binary search. The binary search terminates whenever $\phi_{u} - \phi_{l} = O(\epsilon) \cdot \phi_{s} / 360 = O(\epsilon)\cdot OPT $. Also, we will invoke \Cref{alg:lp} with $\eta = O(\epsilon)$ for an appropriate chosen constant inside $O(\cdot)$. The final approximation ratio is therefore $(3+O(\epsilon))(1+O(\epsilon)) = 3+\epsilon$, for appropriately chosen constants inside the $O(\cdot)$ notations.

Note that in the binary search, $|\phi_{u} - \phi_{l}|$ decreases by half in each iteration. Since it starts at $O(\phi_s)$ and ends at $\Omega(\epsilon \phi_s)$, the binary search takes $O(\log(1/\epsilon))$ rounds. Combining these steps establishes Lemma~\ref{lemma:mainresultmpc}.

\section{Streaming Algorithm}\label{sec:streaming}
\tcbset{
    algbox/.style={
        colframe=black,
        colback=gray!10,
        boxrule=0.8pt,
        arc=3mm,
        left=0mm,
        right=0mm,
        top=0mm,
        bottom=0mm,
        boxsep=0pt,
        sharp corners,
    }
}

In this section, we will show how to adapt Algorithm~\ref{alg:lp} to the streaming model. 

\subsection{Key Challenges}
For the streaming algorithm, we no longer need to compute $\Esim$. To compute $\mathcal{L}$, we first sample $A_v$ for each node at the beginning. Then, for each edge $vx$, we update $A \cdot \vec{N}[x]$ by adding $A_v$ to both $x$ and $v$. We can verify whether $|N[u] \Delta N[v]| \leq_\eta 2\phi$ using Lemma~\ref{lem:test}. The partition for high-degree nodes can then be computed using $O(n \log n / \epsilon^2)$ space.

However, an additional challenge arises, specifically in Lines \ref{ln:R} of Algorithm~\ref{alg:lp}. In Algorithm~\ref{alg:lp}, we select a node $u_i$ and identify all its neighbors that have at most $2\phi$ differing neighbors. While it is straightforward to check whether $|N[w] \Delta N[u_i]| \leq 2\phi$ using the $A$ matrix, determining whether $w \in N(u_i)$ is challenging unless we store all edges incident to $w$.

Fortunately, Assadi and Wang~\cite{AW22} introduced a method that allows sampling each node with probability $\Theta(\log n / d(v))$ and storing all edges incident to the sampled vertices, as stated in the following lemma:

\begin{lemma}[Lemma 3.14 of \cite{AW22}]
\label{lemma:assadi2021}
In $O(n \log n)$ space, with high probability (in the streaming model), we can sample each vertex $v$ with probability $\Theta(\log n / d(v))$ and store all edges incident to the sampled vertices.
\end{lemma}

Using this lemma, we can store the neighbors of a high-degree node whenever there are sufficiently many high-degree nodes inside $L_i$. However, in some cases, there may be very few high-degree nodes in $C^*_i$, and we might only be able to sample a node whose degree is less than $(3 + \eta) \phi$.

\Cref{alg:streamingcp} describes the modifications needed to address this issue. Instead of considering the neighbors of $u_i$, we consider the neighbors of $y_i$, where $y_i$ is any node from $\mathcal{C}^*_{u_i}$. We then check whether these neighbors have at most $2\phi$ differing neighbors with respect to $u_i$ and $y_i$. 

To make the algorithm more general, we also define a candidate set $\cand(L_i)$, which contains all nodes such that the differing neighbors between this node and any node from $L_i$ are at most $2\phi$, while the differing neighbors between this node and any high-degree node not in $L_i$ are greater than $2\phi$. This follows from Theorem~\ref{thm:nostealing}, which implies that any node in $\mathcal{C}^*_{u_i}$ must have a large differing neighborhood with high-degree nodes outside $\mathcal{C}^*_{u_i}$. The modification is shown in Lines~\ref{ln:streamingstart}--\ref{ln:streamingend} of Algorithm~\ref{alg:streamingcp} (within the boxed section).

Note that for any cluster $L_i \in \mathcal{L}$, it contains at least $|L_i|$ nodes, and each node has a degree of at most $|L_i| + \phi \leq 2|L_i|$. Thus, if we sample each node with probability $\Theta(\log n / d(v))$, we can always ensure that we sample some node $y_i \in \mathcal{C}^*_{u_i}$. 
We now show that Algorithm~\ref{alg:streamingcp} remains correct if we modify Lines~\ref{ln:streamingstart}--\ref{ln:streamingend} accordingly. In Section~\ref{sec:streamingimplementation}, we describe how to adapt Algorithm~\ref{alg:streamingcp} to the streaming model.

\begin{algorithm}[t]
\caption{{\textsc{StreamingClusterPhi}}$(G^+=(V, E^+), \phi, \eta)$\label{alg:streaming}\\
\textbf{Input}: A graph $G^{+}$ and parameters $\phi$ and $0 \leq \eta < 1$. \\
\textbf{Output}: A clustering $\mathcal{C}$ with $\obj(\mathcal{C})\leq (3+\eta)\phi$ or ``$\OPT > \phi$''}
\label{alg:streamingcp}
\begin{algorithmic}[1]
\Function {StreamingClusterPhi}{$G^+=(V, E^+), \phi, \eta$}
\Statex {\small \textit{\underline{$\triangleright$ Initialization}}}

    \For{$u,v \in V$} \Comment{$uv$ not necessarily in $E^{+}$.}
        \If{$|N[u] \Delta N[v]| \leq_{\eta} 2 \phi$}
            \State $E' \leftarrow E' \cup \{uv\}$.
        \EndIf
    \EndFor

    \State $\Vlow \leftarrow \{w \mid \mbox{$\deg(w) \leq (3+\eta) \phi$} \}, \Vhigh \leftarrow V \setminus \Vlow$
    \State $V_1 = \Vlow$

    \State Let $\mathcal{L} = \{L_i\}_{i=1}^{|\mathcal{L}|}$ be the partition formed by the connected components in $(\Vhigh ,E')$.

     \For{$i$ from $1$ to $|\mathcal{L}|$} \label{ln:streamingupper}
    \State Choose any node $u_i \in L_i$.
    {
    \begin{tcolorbox}[standard, algbox]
    \State \label{ln:streamingstart} $\cand(L_i) = \{ w \in V_1 \mid |N(u) \Delta N(w) | \leq_{\eta/2} 2\phi \text{ for $u \in L_i$, and } |N(v) \Delta N(w) | \not\leq_{\eta/2} 2\phi \text{ for $v \in \Vhigh \setminus L_i$ }\}$

    \State Let $y_i$ be any vertex in $\mathcal{C}^{*}_{u_i}$.

    \State \label{ln:streamingend} $R(y_i) = \{ w \in V_i \cap \bm{N[y_i]} \mid |N[w] \Delta N[y_i]|  \leq_{\eta/2} 2\phi  \} \cap \cand(L_i)$
    \end{tcolorbox}
    }

 
    \State  \label{ln:streamingC} $C_i \leftarrow L_i \cup R(y_i)$                                       
    \State $V_{i + 1} \leftarrow V_i \setminus R(y_i)$ 

    \EndFor   \label{ln:streaminglower}
     \If{for some $i$ there exists $u \in C_i$ such that $\rho_{C_{i}}(u) > (3+\eta)\phi$}
    \State \Return ``$\OPT > \phi$'' 
    \Else
    \State \Return $\mathcal{C} = \{C_i\}_{i=1}^{|\mathcal{L}|} \cup \bigcup_{v \in V_{|\mathcal{L}|+1}}\{ \{v\} \}$
    \EndIf

\EndFunction
\end{algorithmic}
\end{algorithm}

\subsection{Approximate Ratio}

In this section, we show that the modifications to Lines~\ref{ln:streamingstart}--\ref{ln:streamingend} of Algorithm~\ref{alg:streamingcp} do not affect the approximation ratio. Since we introduce the concept of a candidate set, we first prove that the candidate sets considered in each round are disjoint, ensuring that in each iteration, we consider distinct sets.

\begin{lemma}
\label{lem:canddisjoint}
Assume that $\OPT \leq \phi$. For any $i \neq j$, the candidate sets are disjoint, i.e., $\cand(L_i) \cap \cand(L_j) = \emptyset$. Moreover, we have $\mathcal{C}^*_{u_i} \cap V_1 \subset \cand(L_i)$.
\end{lemma}

\begin{proof}
We proceed by contradiction. Suppose there exists a node $w \in \cand(L_i) \cap \cand(L_j)$. Since $w \in \cand(L_i)$, we know that for any $v \in L_j$, it holds that $|N(v) \Delta N(w)| \not\leq_{\eta / 2} 2\phi$, which contradicts the assumption that $w \in \cand(L_j)$. Hence, the candidate sets must be disjoint.

For any node $w \in \mathcal{C}^*_{u_i} \cap V_1$, Lemma~\ref{lem:highdegreeclustering} ensures that $|N(u) \Delta N(w)| \leq_{\eta/2} 2\phi$ for all $u \in \mathcal{C}^*_{u_i} \cap \Vhigh$. On the other hand, Theorem~\ref{thm:nostealing} states that $|N(w) \Delta N(v)| \not\leq_{\eta/2} 2\phi$ for any high-degree node not in $\mathcal{C}^*_{u_i}$. Thus, $\mathcal{C}^*_{u_i} \cap V_1 \subset \cand(L_i)$, completing the proof.
\end{proof}

We only add nodes from $\cand(L_i)$ as a condition for $w$ to be included in $R(y_i)$. Based on Lemma~\ref{lem:canddisjoint}, we conclude that $R(y_i)$ does not steal low-degree vertices from other clusters in the optimal clustering. 

Now, we are able to state our key lemma. The proof follows a similar structure to the approximation ratio proof for Algorithm~\ref{alg:lp}. The analogues of \Cref{lem:inclusion} and \Cref{lem:closeness} hold as follows.

\begin{lemma}\label{lem:inclusion2} 
Let $C^{*}_i$ be the cluster in $\mathcal{C}^{*}$ such that $L_i \cap \Vhigh = {C}^{*}_{i} \cap \Vhigh$. For any $i$, we have $C^{*}_i \cap \bm{N[y_i]} \subseteq C_i$. 
\end{lemma}

\begin{proof}
If $w \in C^{*}_i \cap \Vhigh$, then $w \in C_i$ because $C_i = L_i \cup R(u_i)$ and ${L}_i \cap \Vhigh = C^{*}_i \cap \Vhigh$ by assumption. 

Now, consider the case where $w \in C^{*}_i \cap \Vlow \cap N[y_i]$. To establish that $w \in C_i$, it suffices to show that $|N[w] \Delta N[y_i]| \leq_{\eta/2} 2\phi$ and $w \in \cand(L_i)$. 

Since $w, y_i \in C^{*}_i$, by \Cref{lem:samecluster}, we have $|N[w] \Delta N[y_i]| \leq 2\phi$, which implies that $|N[w] \Delta N[y_i]| \leq_{\eta/2} 2\phi$ must hold. Furthermore, by \Cref{lem:canddisjoint}, since $w \in C^*_i$, we conclude that $w \in C^*_i \cap \Vlow \subset \cand(L_i)$. 
\end{proof}

Similar to \Cref{lem:closeness}, we show that $N[y_i]$, $C_i$, and $C^*_i$ do not differ a lot.

\begin{lemma}
\label{lem:closeness2} 
Let $C^{*}_i$ be the cluster in $\mathcal{C}^{*}$ such that ${L}_i \cap \Vhigh = {C}^{*}_{i} \cap \Vhigh$. For any $i$, $|N[y_i] \Delta C_i| \leq \phi$ and $|C^{*}_i \Delta C_{i}| \leq \phi$. 
\end{lemma}

\begin{proof}
We first show that $|N[y_i] \Delta C_i| \leq \phi$.
\begin{align}
|\Vlow \cap (N[y_i] \Delta C_i)| &= |\Vlow \cap (N[y_i] \setminus C_i)| + |\Vlow \cap (C_i \setminus N[y_i])| \nonumber \\
&= |\Vlow \cap (N[y_i] \setminus C_i)|  && (\Vlow \cap C_i) \subseteq N[y_i]  \nonumber  \\
&\leq |\Vlow \cap (N[y_i] \setminus C^{*}_i)|  && \mbox{by \Cref{lem:inclusion2}}   \nonumber \\
&\leq |\Vlow \cap (N[y_i] \Delta C^{*}_i)|  \label{eqn:streamingdiffneighborCstar1}
\end{align}

Moreover, since $C^{*}_i \cap \Vhigh = C_i \cap \Vhigh$, it must be the case that 
\begin{align}
|(N[y_i] \Delta C_i) \cap \Vhigh| = |(N[y_i] \Delta C^{*}_i) \cap \Vhigh|  \label{eqn:streamingdiffneighborCstar2}
\end{align}

Therefore,
\begin{align*}
|N[y_i] \Delta C_i| &= |(N[y_i] \Delta C_i) \cap \Vlow| + |(N[y_i] \Delta C_i) \cap \Vhigh| \\
&\leq |(N[y_i] \Delta C^{*}_i) \cap \Vlow| + |(N[y_i] \Delta C^{*}_i) \cap \Vhigh| && \mbox{by (\ref{eqn:streamingdiffneighborCstar1}) and (\ref{eqn:streamingdiffneighborCstar2})} \\
& = |N[y_i] \Delta C^{*}_i| \leq \phi  && \mbox{$u_i \in C^{*}_i$ and $\rho_{C^{*}}(y_i) \leq \phi$}
\end{align*}

Now we show that $|C^{*}_i \Delta C_{i}| \leq \phi$. 

\begin{align*}
&|C^{*}_i \Delta C_{i}| \\
&= |\Vlow \cap (C^{*}_i \Delta C_{i})| && C^{*}_i \cap \Vhigh = C_i \cap \Vhigh \\
&= |\Vlow \cap (C^{*}_i \setminus C_{i})| + |\Vlow \cap (C_i \setminus C^{*}_{i})| \\
&\leq |\Vlow \cap (C^{*}_i \setminus N[y_i])| + |\Vlow \cap (C_i \setminus C^{*}_{i})| && C^{*}_i \cap N[y_i] \subseteq C_i \\
&\leq |\Vlow \cap (C^{*}_i \setminus N[y_i])| + |\Vlow \cap (N[y_i] \setminus C^{*}_{i})| && \Vlow \cap C_i \subseteq V_1 \cap N[y_i] \\
&= |\Vlow \cap (C^{*}_i \Delta N[y_i])| \leq |C^{*}_i \Delta N[y_i]| \leq \phi
\end{align*}

\end{proof}

Using \Cref{lem:closeness2}, we can show that the approximation ratio remains the same for Algorithm~\ref{alg:streamingcp}.

\begin{theorem} Suppose that $\OPT \leq \phi$, the above modification outputs a clustering $\mathcal{C}$ with $\obj(\mathcal{C})\leq (3+\eta) \phi$.  \end{theorem}
\begin{proof}
Consider a component $C_i$ of $\mathcal{C}$. If $C_i$ is a singleton $\{v\}$ with $\deg(v) \leq (3+\eta)\phi$, then obviously, $\rho_{\mathcal{C}}(v) \leq (3+\eta)\phi$. Otherwise, $C_i$ contains some vertex $x$ with $\deg(x) >  (3+\eta)\phi$. By \Cref{lem:highdegreeclustering}, there exists $C^{*}_i$ such that $C_i \cap \Vhigh = {C}^{*}_{i} \cap \Vhigh$.

Let $v$ be any vertex in $C_i$. We will show that $\rho_{\mathcal{C}}(v) \leq 3\phi$. Suppose that $v \in C^{*}_i \cap C_i$, we have: 
\begin{align*}
\rho_{\mathcal{C}}(v) &= |N[v] \Delta C_i|\\
&\leq |N[v] \Delta C^{*}_i | + |C^{*}_i \Delta C_i|  && \mbox{by \Cref{lem:triangleineq}}\\
&= \rho_{\mathcal{C^{*}}}(v) + |C^{*}_i \Delta C_i| && \mbox{by \Cref{lem:closeness2}}\\
&\leq \phi + \phi  = 2\phi
\end{align*}
Otherwise, if $v \notin C^{*}_i \cap C_i$ then it must be the case that $v \in C_i \setminus C^{*}_i$. In such a case, $v$ must be a vertex in $R(u_i)$. This implies $|N[v] \Delta N[y_i]| \leq_{\eta/2} 2\phi$ and thus $|N[v] \Delta N[y_i]| \leq (1+\eta/2)\cdot2\phi$. Therefore:
\begin{align*}
\rho_{\mathcal{C}}(v) = |N[v] \Delta C_i| 
&\leq |N[v] \Delta N[y_i]| + |N[y_i] \Delta C_i| && \mbox{by \Cref{lem:triangleineq} and \Cref{lem:closeness2}} \\
&\leq (1+\eta/2)\cdot 2\phi + \phi = (3+\eta)\phi
\end{align*}
\end{proof}

\subsection{Implementation in the streaming model}
\label{sec:streamingimplementation}

To implement Algorithm~\ref{alg:streamingcp} in the streaming model, we still have several challenges to address. First, the cluster $\mathcal{C}^*_{u_i}$ belongs to the optimal solution, meaning we cannot directly verify whether a sampled node $y_i$ is in $\mathcal{C}^*_{u_i}$. To overcome this, instead of selecting a node from $\mathcal{C}^*_{u_i}$, we consider every node $y$ from the sampled set $S$ that also belongs to the candidate set $\cand(L_i)$:

\[
    R(y) = \{ w \in V_i \cap \bm{N[y]} \mid |N[w] \Delta N[y]|  \leq_{\eta/2} 2\phi \} \cap \cand(L_i).
\]

If $y \in \mathcal{C}^*_{u_i}$, then from the previous section, we know that the final cluster $C_i$ satisfies $\max_{u \in C_i} \rho_{C_i}(u) \leq (3 + \eta) \phi$. If $y \not\in \mathcal{C}^*_{u_i}$, it is still possible to find a cluster $C_i = L_i \cup R(y)$ such that $\max_{u \in C_i} \rho_{C_i}(u) \leq (3 + \eta) \phi$. 

A key observation in this case is that we never steal any low-degree nodes from another cluster $C^*_j$ for $j > i$. This follows from the fact that every time we include $w$ into $R(y)$, we have $w \in \cand(L_i)$, and by Lemma~\ref{lem:canddisjoint}, the candidate sets are disjoint. 

Combining these observations, we conclude that even if we choose an incorrect node $y \notin \mathcal{C}^*_{u_i}$, it does not affect later iterations of the algorithm. Moreover, the order of iteration does not matter—that is, we can process different high-degree node clusters $L_i$ in any order, and the final clustering remains the same when $\OPT \leq \phi$. This property is crucial when using random projection for estimation.

\textbf{Algorithm Description} Now, we describe the streaming implementation, given in Algorithm~\ref{alg:streamingcpfinal}. The algorithm samples a matrix $A$ to check whether $|N[u] \Delta N[v]| \leq_{\eta} 2\phi$, which requires $O(n\log n/\epsilon^2)$ space. Then, using Lemma~\ref{lemma:assadi2021}, it samples each node with probability $\Theta(\log n / d(v))$ along with its neighboring edges. To execute Lines~\ref{ln:streamingfinalstart}--\ref{ln:streamingfinalend}, we iterate over every node in $S \cap \cand(L_i)$. 

We use random projection to compute $\cand(L_i)$, and checking the objective value of $C_i$ can also be done via random projection. At the end, after processing all high-degree clusters, we either output a clustering or report that the current $\OPT > \phi$. 

Throughout the process, we must be careful when using random projection for estimation. This is because dependencies exist both within the same round and across different rounds for $C_i$, making it difficult to bound the probability of correct estimation.

\begin{algorithm}[H]
\caption{{\textsc{StreamingClusterPhiFinal}}$(G^+=(V, E^+), \phi, \eta)$\label{alg:streamingfinal}\\
\textbf{Input}: A graph $G^{+}$ and parameters $\phi$ and $0 \leq \eta < 1$. \\
\textbf{Output}: A clustering $\mathcal{C}$ with $\obj(\mathcal{C})\leq (3+\eta)\phi$ or ``$\OPT > \phi$''}
\label{alg:streamingcpfinal}
\begin{algorithmic}[1]
\Function {StreamingClusterPhiFinal}{$G^+=(V, E^+), \phi, \eta$}
\Statex {\small \textit{\underline{$\triangleright$ Initialization}}}

    \For{$u,v \in V$} \Comment{$uv$ not necessarily in $E^{+}$.}
        \If{$|N[u] \Delta N[v]| \leq_{\eta} 2 \phi$}
            \State $E' \leftarrow E' \cup \{uv\}$.
        \EndIf
    \EndFor

    \State $\Vlow \leftarrow \{w \mid \mbox{$\deg(w) \leq (3+\eta) \phi$} \}, \Vhigh \leftarrow V \setminus \Vlow$
    \State $V_1 = \Vlow$

    \State Let $\mathcal{L} = \{L_i\}_{i=1}^{|\mathcal{L}|}$ be the partition formed by the connected components in $(\Vhigh ,E')$.
        
    \begin{tcolorbox}[standard, algbox]
    
    \State Sample each node $v \in V$ with probability $\Theta(\log n / d(v))$ independently.
    
    \State Let the set of sampled nodes be $S$.

    \end{tcolorbox}

    \For{$i$ from $1$ to $|\mathcal{L}|$} \label{ln:streamingfinalupper}
    \State \label{ln:streamingfinalstart} $\cand(L_i) = \{ w \in V_1 \mid |N(u) \Delta N(w) | \leq_{\eta/2} 2\phi \text{ for $u \in L_i$, and } |N(v) \Delta N(w) | \not\leq_{\eta/2} 2\phi \text{ for $v \in \Vhigh \setminus L_i$ }\}$
    \begin{tcolorbox}[standard, algbox]
    \For{$y \in S \cap \cand(L_i)$}   

    \State \label{ln:streamingfinalend} $R = \{ w \in V_i \cap \bm{N[y]} \mid |N[w] \Delta N[y]|  \leq_{\eta/2} 2\phi  \} \cap \cand(L_i)$
    \State  \label{ln:streamingfinalc}$C_i \leftarrow L_i \cup R$
    \If{$\max_{u \in C_i} \rho_{C_{i}}(u) \leq (3+\eta) \phi$}
        \State Break
    \EndIf
    \EndFor
    \If{$\max_{u \in C_i} \rho_{C_{i}}(u) \leq (3+\eta)\phi$}
        \State $V_{i + 1} \leftarrow V_i \setminus R$
    \Else
        \State \Return ``$\OPT > \phi$''
    \label{ln:streamingfinalend2} \EndIf
    \end{tcolorbox}
    
    \EndFor   \label{ln:streamingfinallower}
\State \Return $\mathcal{C} = \{C_i\}_{i=1}^{|\mathcal{L}|} \cup \bigcup_{v \in V_{|\mathcal{L}|+1}}\{ \{v\} \}$

\EndFunction
\end{algorithmic}
\end{algorithm}

\textbf{Remove dependency for random projection} One technical challenge is checking the value $\max_{u \in C_i} \rho_{C_i}(u)$. The matrix $A$ cannot be used for random projection in this step because it was originally used to compute $C_i$. If we treat $\vec{C_i} \in \{0,1\}^n$ as the characteristic vector of $C_i$ and use $|A \cdot \vec{C_i} - A \cdot \vec{N}[u]|$ to estimate $|C_i \Delta N[u]|$, there exists a dependency between $C_i$ and $A$. Moreover, there may be up to $2^n$ different $C_i$, and if $C_i$ is chosen in a way that maximizes the difference between $|A \cdot \vec{C_i} - A \cdot \vec{N}[u]|$ and $|C_i \Delta N[u]|$, we can no longer apply Lemma~\ref{lem:test} for estimation.

To resolve this, we introduce a second $k \times n$ matrix $B$. When estimating $|C_i \Delta N[u]|$, we compute $|B\cdot \vec{C_i} - B\cdot \vec{N}[u]|$. Since $C_i$ is independent of $B$, random projection yields a correct estimate with high probability.

Another issue arises due to potential dependencies between different rounds of Lines~\ref{ln:streamingfinalstart}--\ref{ln:streamingfinalend}. This dependency is not problematic for the $A$ matrix, as at most $O(n^2)$ neighborhood queries need to be evaluated. Since each query succeeds with high probability, a union bound ensures that all queries for a pair of nodes $u, v$ are answered correctly with high probability. However, arguing about the dependency between different rounds for the $B$ matrix is more delicate. 

Indeed, if we remove the candidate set, adding nodes in the $i$-th iteration may affect later iterations, influencing how $C_i$ is chosen. Fortunately, thanks to the candidate set, Lemma~\ref{lem:canddisjoint} ensures that the nodes considered in each round are disjoint. Thus, $C_i$ does not depend on any $C_j$ for $j \neq i$ and only depends on the randomness of the $A$ matrix and the graph itself. Consequently, when using the $B$ matrix to estimate $|C_i \Delta N[u]|$, it is valid to reuse $B$ across different rounds.

Lastly, we must show the existence of $y \in \mathcal{C}^*_{u_i}$ to guarantee a good cluster. This is formally captured in the following lemma:

\begin{lemma}
Consider Algorithm~\ref{alg:streamingcpfinal}. Assume that $\OPT \leq \phi$. For each $i$, let $C^*_i$ be the cluster in $\mathcal{C}^*$ such that $L_i \cap \Vhigh = C^*_i \cap \Vhigh$. Then, with high probability, $|C^*_i \cap \cand(L_i) \cap S | \geq 1$.
\end{lemma}

\begin{proof}
From Lemma~\ref{lem:canddisjoint}, we know that $C^*_i \cap V_i = C^*_i \cap V_1$, meaning no nodes are lost at iteration $i$ for $C^*_i$. Note that there exists a node $u_i \in C^*_i$ with $d(u_i) \geq 3\phi$. We also know that $d(u_i) - \phi \leq |C^*_i| \leq d(u_i) + \phi$, since otherwise we would have $\rho_{C^*_i}(u_i) > \phi$. Additionally, for any $v \in C^*_i$, we must have $d(v) \leq d(u_i) + 2\phi \leq d(u_i) + (2/3)d(u_i) \leq (5/3) d(u_i)$, as otherwise $\rho_{C^*_i}(v) > \phi$. As $d(v) = O(d(u_i))$, each $v$ is sampled with probability at least $O(\log n / d(u_i))$

Thus, at least $d(u_i) - \phi \geq d(u_i) / 2$ nodes in $C^{*}_i$ are sampled with probability at least $\Omega(\frac{\log n}{d(u_i)})$. By a standard Chernoff bound, we conclude that at least one node from $C^*_i$ is sampled with high probability.
\end{proof}

Combining all points together, we now present our main lemma for the streaming model.

\begin{lemma}[Item \ref{itm:main3} of \Cref{thm:thmmain}]
\label{lemma:mainresultstreaming}
Given a min-max correlation clustering instance $G = (V, E)$, for any $\epsilon > 0$, there exists a randomized streaming algorithm that, in one round, outputs a clustering that is a $(3 + \epsilon)$-approximation for min-max correlation clustering. This algorithm succeeds with high probability and uses a total space of $O(n \log n/\epsilon^2)$.
\end{lemma}

\begin{proof}
We start with $\phi = n$ and run Algorithm~\ref{alg:streamingcpfinal}, which either returns a clustering $\mathcal{C}$ such that $\obj(\mathcal{C}) \leq (3 + \eta) \phi$ or determines that $\OPT > \phi$. To refine $\phi$, we perform a binary search, which takes $O(\log n)$ rounds.

During this process, we need to store the sampled nodes and their neighbors, requiring $O(n\log n)$ space. Additionally, we store the vector $A\cdot \vec{N}[u]$ to answer differing neighbor queries. To compute $\rho_{C_i}(u)$, we use the matrix $B$, where we store $B \cdot \vec{N}[v]$ and compute $B\cdot C_i$ as $\sum_{u} B_u$. 

In summary, Algorithm~\ref{alg:streamingcpfinal} produces a $(3+\epsilon)$-approximate solution with a space complexity of $O(n \log n / \epsilon^2)$.
\end{proof}

\bibliographystyle{alpha}
\bibliography{references}

\appendix

\end{document}